\documentclass[10pt,journal,compsoc]{IEEEtran}
\IEEEoverridecommandlockouts
\usepackage{cite}
\usepackage{amsmath,amssymb,amsfonts}
\usepackage{algorithmic}
\usepackage{graphicx}
\usepackage{textcomp}
\usepackage{xcolor}
\usepackage{silence}
\usepackage[utf8]{inputenc} 
\usepackage[T1]{fontenc}    
\usepackage{hyperref}       
\usepackage{url}            
\usepackage{booktabs}       
\usepackage{nicefrac}       
\usepackage{microtype}      
\usepackage{color}
\usepackage{multirow, multicol}

\usepackage[utf8]{inputenc} 
\usepackage[T1]{fontenc}
\usepackage{amsthm}
\theoremstyle{definition}
\newtheorem{definition}{Definition}[section]

\usepackage{color,soul}

\newcommand{\R}{\mathbb{R}}

\newcommand{\cD}{\mathcal{D}}

\newcommand{\cG}{\mathcal{G}}

\newcommand{\cL}{\mathcal{L}}

\newcommand{\cY}{\mathcal{Y}}

\usepackage[hang,flushmargin]{footmisc}

\def\BibTeX{{\rm B\kern-.05em{\sc i\kern-.025em b}\kern-.08em
    T\kern-.1667em\lower.7ex\hbox{E}\kern-.125emX}}
\begin{document}

\title{Adversarial Attack and Defense on Graph Data: A Survey}

\author{Lichao Sun, 
        Yingtong Dou, 
        Carl Yang, 
        Kai Zhang,
        Ji Wang, 
        Yixin Liu,
        Philip S. Yu,~\IEEEmembership{Fellow,~IEEE,}
        Lifang He, 
        and~Bo~Li, 
        \IEEEcompsocitemizethanks{\IEEEcompsocthanksitem Lichao Sun, Kai Zhang, Yixin Liu, Lifang He are with the Department of Computer Science and Engineering, Lehigh University, Bethlehem, PA, 18015 USA. \protect E-mail: \{lis221, kaz321, yila22, lih319\}@lehigh.edu.
        \IEEEcompsocthanksitem Yingtong Dou, and Philip S. Yu are with the University of Illinois Chicago, Chicago, IL, 60607 USA. \protect E-mail: \{ydou5, psyu\}@uic.edu
        \IEEEcompsocthanksitem Carl Yang is with the Emory University Atlanta, GA 30322 USA. \protect E-mail: j.carlyang@emory.edu
	    \IEEEcompsocthanksitem Ji Wang is with the College of Systems Engineering, National University of Defense Technology, Changsha, Hunan, 410073 P. R. China. \protect E-mail: wangji@nudt.edu.cn.
	    \IEEEcompsocthanksitem Bo Li is with the University of Illinois Urbana-Champaign at Champaign, IL 61820 USA. \protect E-mail: lbo@illinois.edu
	    }
}

\maketitle

\begin{abstract}
    Deep neural networks (DNNs) have been widely applied to various applications, including image classification, text generation, audio recognition, and graph data analysis. However, recent studies have shown that DNNs are vulnerable to adversarial attacks. Though there are several works about adversarial attack and defense strategies on domains such as images and natural language processing, it is still difficult to directly transfer the learned knowledge to graph data due to its representation structure. Given the importance of graph analysis, an increasing number of studies over the past few years have attempted to analyze the robustness of machine learning models on graph data. Nevertheless, existing research considering adversarial behaviors on graph data often focuses on specific types of attacks with certain assumptions. In addition, each work proposes its own mathematical formulation, which makes the comparison among different methods difficult. Therefore, this review is intended to provide an overall landscape of more than $100$ papers on adversarial attack and defense strategies for graph data, and establish a unified formulation encompassing most graph adversarial learning models. Moreover, we also compare different graph attacks and defenses along with their contributions and limitations, as well as summarize the evaluation metrics, datasets and future trends. We hope this survey can help fill the gap in the literature and facilitate further development of this promising new field \footnote{We also have created an online resource to keep track of relevant research on the basis of this survey at \url{https://github.com/safe-graph/graph-adversarial-learning-literature}.}.
    
\end{abstract}



\section{Introduction}
\label{sec01:intro}
Recent years have witnessed significant success achieved by deep neural networks (DNNs) in a variety of domains, ranging
from image recognition~\cite{he2016deep}, natural language processing~\cite{devlin2018bert}, graph data applications \cite{kipf2016semi,velivckovic2017graph,velivckovic2018deep,hamilton2017inductive,wu2019simplifying}, to healthcare analysis \cite{miotto2017deep},
brain circuit modeling \cite{litjens2017a}, and gene mutation functionality \cite{lee2015the}. 
With the superior performance, deep learning has been applied in several safety and security critical tasks such as self driving \cite{bojarski2016end}, malware detection \cite{sun2016sigpid}, identification \cite{sun2017sequential} and anomaly detection \cite{erfani2016high}.
However, the lack of interpretability and robustness of DNNs makes them vulnerable to adversarial attacks. Szegedy et al.~\cite{szegedy2014intriguing} have pointed out the susceptibility of DNNs in image classification. The performance of a well-trained DNN can be significantly degraded by adversarial examples, which are carefully crafted inputs with a small magnitude of perturbations added.
Goodfellow et al.~\cite{goodfellow2014explaining} analyzed this phenomenon and proposed a gradient-based method (FGSM) to generate adversarial image samples. Different adversarial attack strategies are then proposed to demonstrate the vulnerabilities of DNNs in various settings ~\cite{biggio2013evasion,carlini2017towards,xiao2018spatially}.
For instance, black-box adversarial attacks are later explored based on transferability~\cite{liu2016delving, papernot2017practical} and query feedback from DNN models~\cite{bhagoji2017exploring, brendel2017decision}. Some defense and detection methods have also been followed to mitigate such adversarial behaviors~\cite{madry2017towards, samangouei2018defense}, while various adaptive attacks continue to be proposed showing that detection/defense is hard in general~\cite{athalye2018obfuscated, carlini2017adversarial}.

\begin{table*}
\centering
\caption{Attack and Defense works are categorized by GNN or Non-GNN oriented.}
\label{table::attack_catagory}
\begin{tabular}{l|ll}
\toprule
Category & Type & Paper\\ 
\hline
\multirow{4}{*}{Attack Model} & \multirow{2}{*}{GNN}  & \cite{ zugner2018adversarial,dai2018adversarial,bojchevski2019adversarial,chen2018fast,chen2018link,wang2018attack,sun2018data,zugner2019adversarial,wang2019attacking,bose2019generalizable,xu2019topology,chang2020restricted,sun2019node,ma2019attacking}\\

& & \cite{zang2020graph,takahashi2019indirect,chen2020mga,li2020adversarial,fox2019robust,wu2019adversarial, entezari2020all, xi2020graph, zhang2020backdoor, zugner2020adversarial, tang2020adversarial, he2020stealing, wang2020scalable}\\
\cline{2-3}
& \multirow{2}{*}{Non-GNN}  &  \cite{chen2017practical,waniek2018hiding,waniek2018attack,zhou2019attacking,chen2019ga,zhang2019towards,agterberg2019vertex,xuan2019unsupervised,chen2019multiscale,chen2019time,dey2019manipulating,xuan2020adversarial,gaitonde2020adversarial, hou2019alphacyber} \\

& & \cite{dou2020robust,fang2018poisoning, chen2020network}\\
\hline

\multirow{3}{*}{Defense Model} & \multirow{2}{*}{GNN}  & \cite{xu2018characterizing,feng2019graph,zhang2019comparing,sun2019virtual,chen2019can,zhu2019robust,wang2019adversarial,jin2019power,wu2019adversarial,jin2019latent,deng2019batch,zugner2019certifiable,miller2019improving,tang2020transferring}\\

& & \cite{bojchevski2019certifiable,ioannidis2019graphsac,ioannidis2019edge,entezari2020all,wang2019graphdefense,fox2019robust,miller2020topological,ioannidis2020tensor,zhang2020defensevgae, zhang2020gnnguard, xu2020edog, jin2020graph, peng2020robust, zhang2020gcn}\\
\cline{2-3}
& Non-GNN  &  \cite{dai2019adversarial, pezeshkpour2019investigating,agterberg2019vertex,zhou2019adversarial,hou2019alphacyber,jia2020certified,gaitonde2020adversarial,zhang2019towards, dou2020robust, breuer2020friend, yu2019target, logins2020robustness, kenlay2020stability} \\




\bottomrule
\end{tabular}
\end{table*}

Although there are an increasing number of studies on adversarial attack and defense, current research mainly focuses on image, natural language, and speech domains. The investigative effort on graph data is at its infancy, despite the importance of graph data in many real-world applications. For example, in the credit prediction application, an adversary can easily disguise himself by adding a friendship connection with others, which may cause severe consequences \cite{dai2018adversarial}. Compared with non-graph data, the adversarial analysis of graph data presents several unique challenges: 1) Unlike image data with continuous pixel values, the graph structure are discrete valued. It is difficult to design an efficient algorithm that can generate adversarial examples in the discrete space. 2) Adversarial perturbations are designed to be imperceptible to humans in the image domain, so one can force a particular distance function, such as $\ell_p$-norm distance to be small between adversarial and benign instances. However, in graph data, how to define ``imperceptible'' or ``subtle perturbation'' requires further analysis, measurement and investigation.

Given the importance of graph applications in the context of big data and the successful use of graph neural networks (GNNs), the robustness of GNNs has attracted significant interests from both academia and industry. In recent years, many efforts have been made to explore adversarial attacks and defenses for a set of GNN models. The purpose of this paper is to present a comprehensive taxonomy of existing adversarial learning literature on graph data, to develop a framework to unify most existing approaches, and to explore the future tendencies.
All relevant attack and defense studies are listed in Tables \ref{table::attack_summary} and \ref{table::defense_summary}, primarily on the basis of tasks, strategies, baselines, evaluation metrics and datasets. Despite more than 100 papers published in the last three years, there are several challenges remaining unsolved until now, which we contribute to summarize and introduce in this work as follows.

\textbf{Comprehensive Understanding.} To the best of our knowledge, this survey is the first attempt to present an in-depth and comprehensive understanding of the literature about adversarial attack and defense on graph data. 
It has stimulated (and been cited by) various following-up research in this line \cite{zhu2019robust,jin2019power,yadati2019hypergcn,kumar2020adversary,chang2020restricted,jin2020adversarial,chen2020survey}. 
This paper not only provides a broad perspective and guidance for key adversarial attack and defense technologies in the context of GNNs, but also explains many observations related to non-gradient and non-model-based approaches and gives an insight into future directions.

\textbf{Online Updating Resource.}
We created an open-source repository that includes all relevant works and maintained the update on it in the last two years\footnote{\url{https://github.com/safe-graph/graph-adversarial-learning-literature}}. This repository contains links to all relevant papers and
corresponding codes, which makes it easier for researchers to use and track the latest developments, and could serve as a benchmark library in this area. 
However, many of these papers are preprints and reports which give a preview of research results, we will keep tracking them and weaving any updates into the repository accordingly. 
We hope this resource can foster further research on this important topic, and keep shedding light on all facets of future research and development.

\textbf{Unified Problem Definition.}
Though there have been various attack and defense strategies on graph data, there is no unified approach to characterize their relationships and properties; each model seems to be a result of a unique approach.
It is necessary to establish a good basis for easy understanding of existing models and efficient development of future technologies.
In this review, we pioneer to establish a unified formulation and definition to systematically analyze all adversarial attack models on graph data.
Unlike attacks, defenses on graph data often go beyond adversarial learning, for which we provide additional categories based on their unique strategies.

\textbf{Taxonomy of Adversarial Analysis on Graph Data.}
So far there are over a hundred papers that study adversarial analysis on graph data. Compared with image data and text data, the analyses of graph data are more complex due to variations in the graph structure and task. Listing all papers could help but is not intuitive for readers to quickly understand the similarities and discrepancies between different studies. To this end, we summarize existing works based on GNN and Non-GNN methods, aiming to help readers find the most relevant papers easily. We present our taxonomy with more details in Table \ref{table::attack_catagory}.

\textbf{Datasets and Metrics.}
Due to different goals and data used in previous attack and defense works, it is difficult to compare the results of different studies.
Currently, no one could directly answer the question about ``what attack or defense is the best benchmark in this domain?''. The only way to alleviate this is to build a benchmark like other areas \cite{deng2009imagenet,wang2018glue}. Toward this end, we not only develop taxonomies for previous approaches based on different criteria, but also summarize the corresponding datasets and metrics that are frequently used. We hope this study could pave the way for the community to establish a benchmark for future research and practical selection of models in this area.

The rest of this survey is organized as follows: Section~\ref{sec:graph} provides the necessary background information of graph data and common applications.
Section~\ref{sec:attack} presents the unified problem formulation and discusses the existing adversarial attack works on graph data.
Section~\ref{sec:defense} discusses the existing defense works on graph data.
Section~\ref{sec:metric} summarizes the evaluation and attack metrics used in different studies.
Section~\ref{sec:dataset} describes the details of each dataset and summarizes existing works across datasets.
The last section concludes this review.

\begin{table}[htbp]
\centering
\small
\setlength{\tabcolsep}{2.5pt}
\caption{A lookup table of commonly-used notations.}
\begin{tabular}{c|c|c|c}
\toprule
Notation & Description & Notation & Description \\ \hline
$G$ & original graph & $\cL$ & loss function \\ \hline
$\widehat{G}$ & adversarial graph & $f_{\theta}$ & deep learning model \\ \hline
$v$ & node & $\mathcal{Q}$ & distance function \\ \hline
$e$ & edge & $\epsilon$ & cost budget \\ \hline
$c$ & target component & $\Phi$ & perturbation function \\ \hline
$y$ & ground truth label & $\cD$ & dataset \\ \bottomrule
\end{tabular}
\label{tab:notation}
\end{table}
\vspace{-10pt}
\section{Graph} 
\label{sec:graph}

In this section, we first give the notations of graph data, and then introduce the preliminaries about graph types, learning settings, and application tasks. The most frequently used notaions in the paper are summarized in Table \ref{tab:notation}.

\subsection{Notations}
We use $\mathcal{G}=\{G_i\}_{i=1}^N$ to represent a set of graphs, where $N$ is the number of graphs.
Each graph $G_i$ is generally denoted by a set of nodes $V_i=\{v_j^{(i)}\}$ and edges $E_i=\{e_j^{(i)}\}$,
where $e_j^{(i)} = (v_{j,1}^{(i)},v_{j,2}^{(i)}) \in {V_i} \times {V_i}$ is the edge between the nodes $v_{j,1}^{(i)}$ and $v_{j,2}^{(i)}$.
Both nodes and edges can have arbitrarily associated data such as node features, edge weights and edge directions.
According to these properties, graph data can be classified into different types as follows.

\subsection{Types of Graph Data}
\textbf{Dynamic and Static Graphs.} From a {\em temporal perspective}, graph data can be grouped into static graphs and dynamic graphs. 
A graph is dynamic, denoted as $G^{(t)}$, if any of its nodes, edges, node features, or edges features change over time.
On the contrary, a static graph, denoted as $G$,
consists of a fixed set of nodes and edges without changing over time.

A typical example of static graph is the molecular structure of drugs \cite{duvenaud2015}. Once a drug is developed, its molecular structure does not change over time.
Social network \cite{perozzi2014deepwalk} is a good example of dynamic graphs.
As people often add or remove friendship links in their social networks, the graph of relationships and interactions changes over time.
In most existing attack works, the researchers study the attacks on dynamic graphs.

\textbf{Directed and Undirected Graphs.}
The graphs can be divided into directed and undirected graphs according to whether the direction between the initial node and end node is unidirectional or bidirectional.
A directed graph, denoted as $G^{(Dr)}$, has direction information associated with each edge,
where any directed edge $e_1^{(i)} = (v_1^{(i)}, v_2^{(i)}) \neq (v_2^{(i)}, v_1^{(i)}) = e_2^{(i)}$, while an undirected graph has edges made up of unordered pairs of nodes.

Facebook is a classic undirected graph that $A$ is $B$'s friend means $B$ is $A$'s friend too. In contrast to friendships, links of many real networks such as the World Wide Web (WWW), food webs, neural networks, protein interaction networks and many online social networks are directed or asymmetrically weighted. Twitter is a typical example of directed graph, where the directed edge represents the following information from one user to another.


\textbf{Attributed Graph on Edge.} An attributed graph on edge, denoted as $G^{(A_e)}$, has some features associated with each edge, which is denoted by $x(e_j^{(i)}) \in {\R^{{D_{edge}}}}$.

The weighted graph where each edge has a weight, $x(e_j^{(i)}) \in \R$, is a special case of attributed graph on edges.
A traffic flow graph \cite{li2018diffusion} is a typical example of weighted graph where roads are modeled as edges and road conditions are represented by weights of edges.

\textbf{Attributed Graph on Node.} An attributed graph on node, denoted as $G^{(A_n)}$, has some features associated with each node, which is denoted by $x(v_j^{(i)}) \in {\R^{{D_{node}}}}$.

The e-commerce network \cite{eswaran2017zoobp} with different users can be regarded as an example of attributed graph on node where each user is modeled as nodes with some features like demographics and clicking history.

Note that, directed graph and heterogeneous information networks are special cases of {\em attributed graph}, which are widely used to model different applications.

\subsection{Learning Settings on Graph Data}
This section introduces the different machine learning settings used on graph data.
Before introducing the learning settings, we first provide the notations for mathematical formulation.
We associate the target component $c_i$ within
a graph $G^{c_i} \in \cG$ with a corresponding ground truth label $y_i \in \cY = \{1, 2, \ldots, Y\}$. Here $i \in [1, K]$, $K$ represents the total number of target components, and $Y$ is the number of classes being predicted. 
The dataset $\cD^{(ind)}= \{(c_i, G^{c_i}, y_i)\}^{K}_{i = 1}$ is represented by the target graph component, graph containing $c_i$, and the corresponding ground truth label of $c_i$.
For instance, in a node classification task, $c_i$ represents the node to be classified, and $y_i$ denotes its label within $G^{c_i}$.
Based on the features of training and testing processes,
the learning settings can be classified as inductive and transductive learning.

\textbf{Inductive Learning.} It is the most realistic machine learning setting
where the model is trained by labeled examples, and then predicts the labels of examples never seen during training.
Under the supervised inductive learning setting, the classifier $f^{(ind)}\in F^{(ind)} :\cG \rightarrow \cY$ is optimized:
\begin{align*}
    \cL^{(ind)} = \frac{1}{K}\sum^{K}_{i=1} \cL(f^{(ind)}_\theta(c_i, G^{c_i}), y_i),
\end{align*}
where $\cL(\cdot, \cdot)$ is the cross entropy by default, 
and $c_i$ can be node, link or subgraph of its associated graph $G^{c_i}$.
Note that, two or more different instances, $c_1, c_2, \ldots,$ and $c_n$ can be associated with the same graph $G \in \cG$.

\textbf{Transductive Learning.}
Different from inductive learning, the testing graphs have been seen during training in the transductive learning.
In this case, the classifier $f^{(tra)}\in F^{(tra)} :\cG \rightarrow \cY$ is optimized:
\begin{equation*}
    \cL^{(tra)} = \frac{1}{K} \sum^{K}_{i=1} \cL(f^{(tra)}_\theta(c_{i}, G^{c_i}), y_i).
\end{equation*}
Transductive learning predicts the label of {\em seen} instances,
but inductive learning predicts the label of {\em unseen} instances.

\textbf{Unified Formulation of Learning on Graph Data.}
We give an uniform formula to represent both supervised inductive and transductive learning as below:
\begin{equation}\label{eq:uniformloss}
    \cL^{(\cdot)} = \frac{1}{K} \sum^{K}_{i=1} \cL(f^{(\cdot)}_\theta(c_{i}, G^{c_i}), y_i),
\end{equation}
where $f^{(\cdot)}_\theta = f^{(ind)}_\theta$ is inductive learning and $f^{(\cdot)} = f^{(tra)}_\theta$ is transductive learning.

In the unsupervised learning setting, we can use the unlabelled dataset $\cD^{(ind)}= \{(c_{i}, G_j)\}^{K}_{i = 1}$ and replace the supervised loss $\cL$ and function $f(c_{i}, G_i)$ of Eq.~(\ref{eq:uniformloss}).

In this survey, we mainly focus on the supervised learning setting, while also introducing a few new works in the unsupervised learning setting.

\subsection{Application}

In this section, we will introduce the main tasks on graph data,
including node-level, link-level and graph-level applications.
Moreover, we also introduce how to use the unified formulation of Eq.~(\ref{eq:uniformloss}) to define each application task below.

\textbf{Node-Level Application.}
The node-level application is the most popular one in both academia and industry.
A classic example is labeling the nodes in the Web and social network graphs, 
which may contain millions of nodes, such as Facebook and Twitter.

Most existing papers \cite{bojchevski2019adversarial, dai2018adversarial, zugner2018adversarial,zugner2019adversarial,zugner2019certifiable,bojchevski2019certifiable,zhu2019robust,wang2019attacking,wu2019adversarial,xu2019topology}
focus on node-level applications.
All of these papers study node classification in the transductive learning setting whose objective function can be formulated by modifying Eq.~(\ref{eq:uniformloss})
where $f^{(\cdot)}_\theta = f^{(tra)}_\theta$, $c_i$ here is the representation of node target and its associated graph $G^{c_i}$ is set as a single graph $G$.

Few existing works have discussed the node-level applications in the inductive leaning setting. However, these applications frequently appear in real life.
For example, the first party only has several large and public network information, such as Facebook and Twitter.
The second party has private unlabeled graph data in which the nodes can be predicted by using the information from the first party.
In this case, the node-level classification task is no longer transductive learning. It can be easily formulated by modifying Eq.~(\ref{eq:uniformloss}) with $f^{(\cdot)}_\theta = f^{(ind)}_\theta$ and $c_i$ here is still the representation of node target.

\textbf{Link-Level Application.}
Link prediction on dynamic graphs is one of the most common link-level applications.
The models try to predict missing links in current networks, as well as new or dissoluted links in future networks.
The corresponding attacks and defenses have been discussed in~\cite{sun2018data, zhou2019attacking}.

Compared with node classification tasks, link predication tasks still use node features,
but target at the missing or unlabelled links in the graph.
Therefore, we can formulate the link predication task by slightly modifying Eq.~(\ref{eq:uniformloss})
with $c_{i}$ being the representation of link target, and $y_i \in \{0, 1\}$.

\textbf{Graph-Level Application.} 
Graph-level tasks are frequently seen in the chemistry or medical areas,
such as the modeling of drug molecule graphs and brain graphs. In \cite{dai2018adversarial}, the whole graph is used as the sample instance. Different from this setting, some other graph-level applications use the subgraphs of a larger graph for particular tasks~\cite{xi2020graph,zhang2020backdoor}.

Compared with the existing works on node classification and link predication,
graph classification uses the graph-structure representation as the features to classify the unlabelled graph instances.
Therefore, we can formulate the graph classification task by slightly modifying Eq.~(\ref{eq:uniformloss})
by setting $c_{i}$ as the representation of graph target.


\begin{table*}[hbt!]
\centering
\caption{Summary of adversarial attack works on graph data (time ascending).}
\label{table::attack_summary}
\resizebox{\linewidth}{!}{%
\begin{tabular}{ccccccccccc}
\toprule
Task &Ref. & Year & Venue  &  Model  &  Strategy  &  Approach  & Baseline & Metric & Dataset\\ 
\hline

Graph clustering &\cite{chen2017practical} & 2017 & CCS  & \begin{tabular}[c]{@{}c@{}c@{}}SVD, Node2vec, \\ Community \\ detection algs\end{tabular} & \begin{tabular}[c]{@{}c@{}c@{}}Noise injection, \\ Small community \\ attack\end{tabular}  & Add/Delete edges  & -  & ASR, FPR & \begin{tabular}[c]{@{}c@{}c@{}}NXDOMAIN, \\ Reverse Engineered \\ DGA Domains\end{tabular}\\ 
\hline

\multirow{45}{*}{Node classification} 
& \cite{zugner2018adversarial} & 2018 & KDD &  \begin{tabular}[c]{@{}c@{}}GCN, CLN,\\DeepWalk\end{tabular} & Incremental attack &\begin{tabular}[c]{@{}c@{}}Add/Delete edges, \\ Modify node features\end{tabular} & \begin{tabular}[c]{@{}c@{}}Random,\\FGSM\end{tabular} & \begin{tabular}[c]{@{}c@{}c@{}}Accuracy, \\Classifcation \\margin \end{tabular} &\begin{tabular}[c]{@{}c@{}c@{}}Cora-ML, \\Citeseer,  \\ PolBlogs\end{tabular} \\ \cline{2-10}

& \cite{wang2018attack}& 2018 & arXiv  &   GCN  &  Greedy, GAN  &  \begin{tabular}[c]{@{}c@{}}Add fake nodes \\ with fake features \end{tabular}  &  Random, Nettack  &  \begin{tabular}[c]{@{}c@{}}Accuracy, \\ F1, ASR\end{tabular}  & \begin{tabular}[c]{@{}c@{}}Cora, \\ Citeseer\end{tabular} \\ \cline{2-10}

& \cite{wang2019attacking}& 2019 & CCS  &  \begin{tabular}[c]{@{}c@{}c@{}c@{}}LinBP, LBP, JW,  \\ DeepWalk, LINE, \\GCN, RW, \\Node2vec\end{tabular}   &  Optimization   &  Add/Delete edges   &  Random, Nettack  &  FNR, FPR  & \begin{tabular}[c]{@{}c@{}c@{}}Google+, \\Epinions, Twitter,  \\ Facebook, Enron\end{tabular}\\ \cline{2-10}

& \cite{xu2019topology}& 2019 & IJCAI  &  GCN   &  \begin{tabular}[c]{@{}c@{}}First-order \\ optimization \end{tabular}  &   Add/Delete edges  &  \begin{tabular}[c]{@{}c@{}}DICE, Greedy, \\ Meta-self \end{tabular} &   \begin{tabular}[c]{@{}c@{}}Misclassification\\ rate \end{tabular}  & \begin{tabular}[c]{@{}c@{}}Cora,\\ Citeseer\end{tabular} \\ \cline{2-10}

& \cite{chang2020restricted}& 2019 & AAAI  &  \begin{tabular}[c]{@{}c@{}}GCN, LINE,\\ SGC, DeepWalk\end{tabular}   &  \begin{tabular}[c]{@{}c@{}c@{}} Approximate \\spectrum,
\\ Devise new loss\end{tabular}   &  Add/Delete edges   &  \begin{tabular}[c]{@{}c@{}}Random, Degree,\\ RL-S2V, \end{tabular}  & Accuracy   & \begin{tabular}[c]{@{}c@{}}Cora, Citeseer, \\ Pubmed\end{tabular}\\ \cline{2-10}

& \cite{zugner2019adversarial}& 2019 & ICLR    &  \begin{tabular}[c]{@{}c@{}}GCN, CLN \\ DeepWalk\end{tabular}   &  Meta learning   &  Add/Delete edges &  \begin{tabular}[c]{@{}c@{}}DICE, Nettack,\\ First-order attack \end{tabular} & \begin{tabular}[c]{@{}c@{}c@{}}Accuracy, \\Misclassification\\ rate \end{tabular} & \begin{tabular}[c]{@{}c@{}c@{}}Cora, Pubmed,\\ Citeseer,  \\ PolBlogs\end{tabular}\\ \cline{2-10}

& \cite{ma2019attacking}& 2019 & arXiv   &   GCN  &   \begin{tabular}[c]{@{}c@{}}Reinforcement \\learning\end{tabular}  &  Rewire edges   &  RL-S2V, Random  &  ASR  & \begin{tabular}[c]{@{}c@{}}Reddit-Multi,\\ IMDB-Multi\end{tabular}\\ \cline{2-10}

& \cite{bose2019generalizable}& 2019 & arXiv  &   GCN  &   \begin{tabular}[c]{@{}c@{}}Adversarial \\generation \end{tabular}  & Modify node features  &  Nettack  &  ASR  & \begin{tabular}[c]{@{}c@{}}Cora, \\Citeseer\end{tabular} \\ \cline{2-10}

& \cite{wu2019adversarial} & 2019 & IJCAI &  GCN  &  Check gradients   &   \begin{tabular}[c]{@{}c@{}}Add/Delete edges, \\ Modify node features\end{tabular}  &  \begin{tabular}[c]{@{}c@{}c@{}}Random, \\Nettack \\ FGSM, JSMA\end{tabular}  & \begin{tabular}[c]{@{}c@{}c@{}}Accuracy, \\ Classification \\ margin\end{tabular}   & \begin{tabular}[c]{@{}c@{}}Cora, Citeseer,\\PolBlogs\end{tabular}\\ \cline{2-10}


& \cite{takahashi2019indirect} & 2020 & BigData  &  GCN & Check gradients   &  Modify node features   &  Nettack  &  ASR  &\begin{tabular}[c]{@{}c@{}} Cora-ML, \\Citeseer \end{tabular} \\ \cline{2-10}

& \cite{entezari2020all} & 2020 & WSDM &  GCN, t-PINE   &  \begin{tabular}[c]{@{}c@{}}Low-rank \\approximation   \end{tabular}&     Add/Delete edges 	  &  Nettack  & \begin{tabular}[c]{@{}c@{}c@{}} Correct \\classification\\ rate \end{tabular} & \begin{tabular}[c]{@{}c@{}c@{}} Cora-ML, \\ Citeseer, \\PolBlogs \end{tabular}\\ \cline{2-10}

& \cite{zugner2020adversarial} & 2020 & TKDD  &  \begin{tabular}[c]{@{}c@{}}GCN, CLN, \\ DeepWalk \end{tabular}&  Incremental attack &     \begin{tabular}[c]{@{}c@{}}Add/Delete edges, \\ Modify node features\end{tabular}  & \begin{tabular}[c]{@{}c@{}}Random,\\FGSM\end{tabular} & \begin{tabular}[c]{@{}c@{}c@{}}Accuracy, \\Classifcation \\margin \end{tabular}& \begin{tabular}[c]{@{}c@{}c@{}}Cora-ML, \\Citeseer,  \\ PolBlogs, Pubmed\end{tabular}  \\ \cline{2-10}

& \cite{sun2019node}& 2020 & WWW  &  GCN   &  \begin{tabular}[c]{@{}c@{}}Reinforcement \\learning\end{tabular}  &  Inject new nodes   &  \begin{tabular}[c]{@{}c@{}c@{}}Random, FGA,\\ Preferential\\attack\end{tabular}  & \begin{tabular}[c]{@{}c@{}c@{}}Accuracy,\\ Graph \\ statistics\end{tabular}   & \begin{tabular}[c]{@{}c@{}c@{}}Cora-ML,\\ Pubmed,\\Citeseer\end{tabular}\\ \cline{2-10}

&\cite{ma2020towards} &2020 &NIPS & GCN, JK-Net & Check gradients & Modify node features & \begin{tabular}[c]{@{}c@{}c@{}}Degree, Betweenness,\\ PageRank, Random \end{tabular} & Mis-classification rate & \begin{tabular}[c]{@{}c@{}c@{}} Cora, Citeseer\\ Pubmed \end{tabular} \\ \cline{2-10}

&\cite{geisler2021robustness} &2021 & NIPS &\begin{tabular}[c]{@{}c@{}c@{}} GCN family models, \\GDC, SGC \end{tabular} & Check gradients &Add/delete edges &FGSM, PGD, Acc. &ASR &\begin{tabular}[c]{@{}c@{}c@{}} Cora ML, Citeseer\\ Pubmed, arXiv \\ Products, Paper 100M \\ \end{tabular} \\\cline{2-10}

&\cite{tao2021single} &2021 &CIKM &\begin{tabular}[c]{@{}c@{}c@{}} GCN, GAT\\ APPNP \\ \end{tabular} &Optimization & Inject new nodes &\begin{tabular}[c]{@{}c@{}c@{}} Radnom, MostAttr\\ PrefEdge, NIPA \\ AFGSM, G-NIA \end{tabular} & \begin{tabular}[c]{@{}c@{}c@{}} Misclassifcation rate \end{tabular} &\begin{tabular}[c]{@{}c@{}c@{}} Reddit, Citeseer\\ ogbn-products \\ \end{tabular} \\ \cline{2-10}

&\cite{zou2021tdgia} &2021 &KDD &GCN &Optimization & Inject new nodes/edges &GSM, AFGSM, SPEIT & Classification Accuracy & \begin{tabular}[c]{@{}c@{}c@{}} KDD-CUP, Reddit\\ ogbn-arxiv \end{tabular} \\ \cline{2-10}

& \cite{zang2020graph}& 2021 & IJCAI &  \begin{tabular}[c]{@{}c@{}}GCN, DeepWalk, \\Node2vec, GAT \end{tabular}   &  Check gradients   &  Add/Delete edges   &  \begin{tabular}[c]{@{}c@{}} Random, FGA, \\Victim-class attack \end{tabular}  &   ASR, AML & \begin{tabular}[c]{@{}c@{}c@{}} Cora,\\ Citeseer, \\PolBlogs \end{tabular}\\
\hline

\multirow{12}{*}{Link prediction}

& \cite{sun2018data}& 2018 & arXiv &  \begin{tabular}[c]{@{}c@{}}GAE, DeepWalk,\\ Node2vec, LINE\end{tabular}  &  \begin{tabular}[c]{@{}c@{}}Project\\
 gradient descent\end{tabular}  & Add/Delete edges &  \begin{tabular}[c]{@{}c@{}c@{}} Degree sum,\\Shortest path, \\ Random, PageRank \end{tabular} & \begin{tabular}[c]{@{}c@{}c@{}} AP,\\Similarity\\ score\end{tabular} & \begin{tabular}[c]{@{}c@{}c@{}}Cora, \\ Citeseer,  \\ Facebook\end{tabular}\\ \cline{2-10}
 
& \cite{zhou2019attacking}& 2019 & AAMAS &  \begin{tabular}[c]{@{}c@{}c@{}}Local\&Global \\ Similarity \\measures \end{tabular}  &  Submodular   &   Hide edges  &  Random, Greedy  &  \begin{tabular}[c]{@{}c@{}} Similarity \\score \end{tabular} & \begin{tabular}[c]{@{}c@{}}Random, \\ Facebook \end{tabular}\\ \cline{2-10}

& \cite{chen2019time}& 2021 & TKDE  &  \begin{tabular}[c]{@{}c@{}c@{}}Deep dynamic\\ network\\ embedding algs\end{tabular}   &   Check gradients  &  Rewire edges   &  \begin{tabular}[c]{@{}c@{}}Random, Gradient,\\ Common neighbor\end{tabular}  &  ASR, AML  & \begin{tabular}[c]{@{}c@{}}LKML, FB-WOSN,\\ RADOSLAW\end{tabular}\\ \cline{2-10}

&\cite{bhardwaj2021adversarial} & 2021 & EMNLP &\begin{tabular}[c]{@{}c@{}c@{}} TransE, DistMult\\ ConvE, ComplEx \end{tabular} &Instance attribution &Add/Delete facts &\begin{tabular}[c]{@{}c@{}c@{}}Direct-Add/Del, CRIAGE,\\ Random edits,\\ Gradient Rollback \end{tabular} & \begin{tabular}[c]{@{}c@{}c@{}}MRR\\ Hits@K \end{tabular} & \begin{tabular}[c]{@{}c@{}c@{}} WN18RR\\ FB15k-237 \end{tabular}\\ \cline{2-10}

&\cite{bhardwaj2021poisoning} &2021 & ACL & \begin{tabular}[c]{@{}c@{}c@{}} TransE, DistMult\\ ConvE, ComplEx \end{tabular} &\begin{tabular}[c]{@{}c@{}c@{}} Exploit relation\\ inference patterns \end{tabular} &Create decoy facts &\begin{tabular}[c]{@{}c@{}c@{}} Random, CRIAGE\\ Edits in the neiborhood \end{tabular} &\begin{tabular}[c]{@{}c@{}c@{}} MRR\\ Hits@K \end{tabular} & \begin{tabular}[c]{@{}c@{}c@{}} WN18RR\\ FB15k-237 \end{tabular} \\ \hline

\multirow{4}{*}{Graph classification}



&\cite{wan2021adversarial} & 2021 & NIPS &\begin{tabular}[c]{@{}c@{}c@{}} GCN, GIN\\ Cheby-GIN \\ Graph U-net \\ \end{tabular} & Beyasian optimization &  \begin{tabular}[c]{@{}c@{}c@{}} Add/delete edges\\ Rewire edges \\ Inject new nodes \\ \end{tabular} & \begin{tabular}[c]{@{}c@{}c@{}} Random, Genetic\\ Gradient-based \end{tabular} &ASR & \begin{tabular}[c]{@{}c@{}c@{}} IMDB-M, Proteins\\ Collab, Twitter fake news  \\ Reddict-Multi-5k \\ \end{tabular} \\ \cline{2-10}

&\cite{mu2021hard} & 2021 & CCS & \begin{tabular}[c]{@{}c@{}c@{}} GIN, SAG \\ GUNet \end{tabular} & Optimization & Add/delete edges &Random, RL-S2V &\begin{tabular}[c]{@{}c@{}c@{}} ASR, Average Purturbation, \\Average Queries, Average Time  \end{tabular} & \begin{tabular}[c]{@{}c@{}c@{}} COIL, IMDB, NCI1 \end{tabular} \\ \cline{2-10}

&\cite{zhang2021projective} &2021 & CIKM  & \begin{tabular}[c]{@{}c@{}c@{}} GCN \end{tabular} &Project ranking of elements & Add edges & \begin{tabular}[c]{@{}c@{}c@{}} RandomSampling \\ GradArgmax \\RL-S2V \end{tabular} &\begin{tabular}[c]{@{}c@{}c@{}} Correct \\ classification rate \end{tabular} & \begin{tabular}[c]{@{}c@{}c@{}} BA-2Motifs, ENZYMES, \\Mutagenicity, PC-3, \\NCI109, NCI-H23H \end{tabular}\\ \hline

\multirow{1}{*}{Community detection}


& \cite{chen2019ga}& 2019 & TCSS &  \begin{tabular}[c]{@{}c@{}}Community \\detection algs \end{tabular}  &  Genetic algs   &  Rewire edges   &  \begin{tabular}[c]{@{}c@{}}Random, Degree, \\ Community detection\end{tabular}  &  \begin{tabular}[c]{@{}c@{}} NMI, \\Modularity\end{tabular}  & \begin{tabular}[c]{@{}c@{}c@{}}Karate, Dolphin,  \\ Football, \\ Polbooks\end{tabular}\\ 
\cline{2-10}

& \cite{li2020adversarial}& 2020 & WWW  &  \begin{tabular}[c]{@{}c@{}c@{}}Surrogate\\ community\\ detection model\end{tabular}    & \begin{tabular}[c]{@{}c@{}c@{}}  Graph \\auto-encoder  \end{tabular} & Add/Delete edges    &  \begin{tabular}[c]{@{}c@{}c@{}}DICE, Random, \\ Modularity \\based attack  \end{tabular}&  \begin{tabular}[c]{@{}c@{}}Personalized \\metric \end{tabular}  &  \begin{tabular}[c]{@{}c@{}}DBLP,\\ Finance\end{tabular}\\ \hline


\multirow{4}{*}{\begin{tabular}[c]{@{}c@{}}Node classification,\\ Link prediction\end{tabular}} 

&\cite{bojchevski2019adversarial}& 2019 & ICML  & \begin{tabular}[c]{@{}c@{}}Node2vec, GCN\\ LP, DeepWalk\end{tabular}  & \begin{tabular}[c]{@{}c@{}c@{}} Check gradient, \\ Approximate \\spectrum \end{tabular}  & Add/Delete edges  & \begin{tabular}[c]{@{}c@{}}Random, Degree,\\ Eigenvalue\end{tabular}  & \begin{tabular}[c]{@{}c@{}c@{}}F1 score, \\ Misclassification \\rate \end{tabular}  &  \begin{tabular}[c]{@{}c@{}}Cora, Citeseer,  \\ PolBlogs\end{tabular}  \\ \cline{2-10}

&\cite{gupta2021adversarial} & 2021 & PAKDD  &\begin{tabular}[c]{@{}c@{}c@{}} DeepWalk, Node2Vec,\\ LINE, GCN \end{tabular} &Optimization &add/delete edges & Random, UNSUP & Micro F1, Precision &\begin{tabular}[c]{@{}c@{}c@{}} LFR, Cora, Citeseer\\ ForestFire, PolBlogs \end{tabular}\\ \hline

\begin{tabular}[c]{@{}c@{}}Graph classification,\\ Node classification\end{tabular} &\cite{dai2018adversarial} & 2018 & ICML  & \begin{tabular}[c]{@{}c@{}}GNN family\\ models\end{tabular} & \begin{tabular}[c]{@{}c@{}}Reinforcement\\ learning\end{tabular} &   Add/Delete edges  & \begin{tabular}[c]{@{}c@{}}Rnd. sampling,\\Genetic algs.\end{tabular}  & Accuracy &  \begin{tabular}[c]{@{}c@{}}Citeseer,Finance,  \\ Pubmed, Cora\end{tabular}  \\ \hline


\begin{tabular}[c]{@{}c@{}}Malware detection, \\ Node classification \end{tabular} &\cite{hou2019alphacyber} & 2019 & CIKM  &  
Metapath2vec  &  Greedy   &  Inject new nodes  & Anonymous attack  & \begin{tabular}[c]{@{}c@{}}\%TPR,
 \\ TP-FP curve \end{tabular} & Private dataset\\ 
\hline

\begin{tabular}[c]{@{}c@{}c@{}}Knowledge graph \\fact plausibility\\prediction\end{tabular} &\cite{zhang2019towards}& 2019 & IJCAI  &  \begin{tabular}[c]{@{}c@{}}RESCAL, \\ TransE, TransR\end{tabular}   &   \begin{tabular}[c]{@{}c@{}}Check target \\ entity embeddings\end{tabular}   & Add/Delete fact   &  Random  &   \begin{tabular}[c]{@{}c@{}}MRR, \\Hit Rate$@$K \end{tabular} & FB15k, WN18\\ 
\hline

Vertex nomination &\cite{agterberg2019vertex}& 2019 & arXiv &   VN·GMM·ASE  &   Random  &  Add/Delete edges   &  -  &  \begin{tabular}[c]{@{}c@{}}Achieving \\rank \end{tabular} & \begin{tabular}[c]{@{}c@{}}Bing entity \\transition graph\end{tabular}\\ 
\hline

\begin{tabular}[c]{@{}c@{}} Manipulating \\ opinion \end{tabular} &\cite{gaitonde2020adversarial}& 2020 & arXiv  &  Graph model &  \begin{tabular}[c]{@{}c@{}} Adversarial\\ optimization \end{tabular} &  \begin{tabular}[c]{@{}c@{}} Change initial\\opinion vector \end{tabular} &  -  &  -  & -\\
\hline

Fraud detection &\cite{dou2020robust} & 2020 & KDD  &  \begin{tabular}[c]{@{}c@{}}Graph-based\\ Fraud detectors\end{tabular}  &  \begin{tabular}[c]{@{}c@{}} Reinforcement \\ learning\end{tabular}  &  Add/Delete edges   &   -   &  \begin{tabular}[c]{@{}c@{}}
Practical\\ effect\end{tabular}  & \begin{tabular}[c]{@{}c@{}c@{}}YelpChi, \\YelpNYC,\\ YelpZip\end{tabular}\\ 
\hline

Graph matching &\cite{zhang2020adversarial} & 2020 & NIPS & \begin{tabular}[c]{@{}c@{}c@{}}SNNA, DGMC \\ CrossMNA \end{tabular} & \begin{tabular}[c]{@{}c@{}c@{}}Kernel density estimation, \\ Meta learning \end{tabular} & Inject new nodes & \begin{tabular}[c]{@{}c@{}c@{}} Random, RL-S2V \\ Meta-Self, CW-PGD \\ GF-Attack, CD-ATTACK \end{tabular} & \begin{tabular}[c]{@{}c@{}c@{}}Accuracy, \\ Precision@K \end{tabular} & \begin{tabular}[c]{@{}c@{}c@{}}Autonomous systems \\ Last.FM, DBLP \\ LiveJournal \end{tabular}\\
\hline

\begin{tabular}[c]{@{}c@{}c@{}} Knowledge graph\\ alignment \end{tabular} &\cite{zhang2021adversarial} & 2021 & EMNLP  &GCN &\begin{tabular}[c]{@{}c@{}c@{}} Kernel density\\ estimation \end{tabular} &Add/Delete relations & \begin{tabular}[c]{@{}c@{}c@{}} SWS, IWS, DPA\\ GF-Attack, LowBlow \\ CRIAGE, RL-RR \end{tabular} & \begin{tabular}[c]{@{}c@{}c@{}}MRR\\ His@K \end{tabular} & DBP15K \\
\hline

\begin{tabular}[c]{@{}c@{}c@{}}Question answering, \\ Item recommendation \end{tabular} &\cite{raman2020learning} & 2021 & ICLR &\begin{tabular}[c]{@{}c@{}c@{}}RN, MHGRN \\ KGCN, RippleNet \end{tabular} & \begin{tabular}[c]{@{}c@{}c@{}}Reinforcement learning, \\ Heuristic \end{tabular} &\begin{tabular}[c]{@{}c@{}c@{}}Replace relations \end{tabular} &Random & \begin{tabular}[c]{@{}c@{}c@{}} Accuracy, AUC, \\Aggregated triple score, \\ Similarity in clustering coefficient\\ /degree distribution \end{tabular} & \begin{tabular}[c]{@{}c@{}c@{}} CSQA, OBQA \\Last.FM \\ MovieLens-20M \end{tabular} \\
\hline

Malware detection &\cite{zhao2021structural} & 2021 & CCS & FCG & \begin{tabular}[c]{@{}c@{}c@{}}Heuristic optimization, \\ Reinforcement learning \end{tabular} & \begin{tabular}[c]{@{}c@{}c@{}}Add/Rewire edges \\ Insert/Delete nodes \end{tabular} & - & \begin{tabular}[c]{@{}c@{}c@{}}Initialization/Relative/ \\ /Absolute ASR \end{tabular} & Malscan\\ \hline

Node Similarity &\cite{dey2019manipulating}& 2020 & AAMAS  &  \begin{tabular}[c]{@{}c@{}}Similarity \\measures\end{tabular} &  Graph theory   &  Remove edges   & \begin{tabular}[c]{@{}c@{}c@{}}Greedy, Random, \\High jaccard \\similarity\end{tabular}   &   \begin{tabular}[c]{@{}c@{}}\# Removed \\edges\end{tabular}  & \begin{tabular}[c]{@{}c@{}c@{}}Power,web-edu,\\ hamsterster,\\euroroad\end{tabular}\\ 

\bottomrule
\end{tabular}
}
\end{table*}

\section{Adversarial Attacks on Graph Data} \label{sec:attack}

In this section, we give a general definition and taxonomies of adversarial attacks on graph data,
and then introduce the imperceptibility metrics, attack types, attack tasks and levels of attack knowledge.

\subsection{An Unified Definition and Formulation}

\begin{definition}(General Adversarial Attack on Graph Data)
    Given a dataset $\cD = (c_{i}, G_i, y_i)$, after slightly modifying $G_i$ (denoted as $\widehat{G}^{c_i}$),
    the adversarial samples $\widehat{G}^{c_i}$ and $G_i$ should be similar under the imperceptibility metrics,
    but the performance of graph task becomes much worse than before.
\end{definition}

Existing papers~\cite{chen2017practical,bojchevski2019adversarial, chen2018link, dai2018adversarial, sun2018data, zugner2018adversarial,zugner2019adversarial,wang2019attacking,wu2019adversarial,xu2019topology,chang2020restricted,hou2019alphacyber,li2020adversarial,entezari2020all} considering adversarial behaviors on graph data usually focus on specific types of attacks with certain assumptions.
In addition, each work proposes its own mathematical formulation which makes the comparison among different methods difficult.
In order to help researchers understand the relations between different problems,
we propose a unified problem formulation that can cover all current existing works.

\begin{definition}(Adversarial Attack on Graph Data: A Unified Formulation)
    $f$ can be any learning task function on graph data,
    e.g., link prediction, node-level embedding, node-level classification,
    graph-level embedding and graph-level classification.
    ${\Phi(G_i)}$ denotes the space of perturbation on the original graph $G_i$,
    and dataset $\widehat{\cD}$ = $\{(c_{i}, \widehat{G}^{c_i}, y_i)\}_{i=1}^N$ denote the attacked instances.
    The attack can be depicted as,
    \begin{equation}\label{eq:uniformattack}
    \begin{aligned}
    & \underset{\widehat{G}^{c_i} \in \Phi(G_i)}{\text{max}}
    & & \sum_i \cL(f^{(\cdot)}_{\theta^*}(c_{i}, \widehat{G}^{c_i}), y_i)) \\
    & \text{s.t.}
    & & {\theta ^*} = \mathop {\arg \min }\limits_\theta 
    \sum_j {\cL}({f^{(\cdot)}_\theta }(c_{j}, G'_j), y_j)).
    \end{aligned}
    \end{equation}
When $G'_j$ equals to $\widehat{G}^{c_j}$, Eq.~(\ref{eq:uniformattack}) represents the poisoning attack, whereas when $G'_j$ is the original $G$ without modification, Eq.~(\ref{eq:uniformattack}) denotes the evasion attack.
$f^{(\cdot)}_\theta = f^{(ind)}_\theta$ represents inductive learning and $f^{(\cdot)}_\theta = f^{(tra)}_\theta$ transductive learning.
\end{definition}
Note that, with $\widehat{G}^{c_i} \in \Phi(G)$, $(c_i, \widehat{G}^{c_i})$ can represent node manipulation, edge manipulation, or both.
For any $\widehat{G}^{c_i} \in \Phi(G_i)$, $\widehat{G}^{c_i}$ is required to be similar or close to the original graph $G_j$, and such similarity measurement can be defined by the general distance function below:
\begin{equation}\label{eq:distribution}
    \begin{aligned}
    & & \mathcal{Q}( \widehat{G}^{c_i}, G_i) < \epsilon \\
    & \text{s.t.}
    & \widehat{G}^{c_i} \in \Phi(G_i)
    \end{aligned}
\end{equation}
where $\mathcal{Q}(\cdot, \cdot)$ represents the distance function, and $\epsilon$ is a parameter denoting the distance/cost budget for each sample.

\textbf{Discussion: Graph Distance Function.}
Graph distance functions can be defined in many ways,
a lot of which have been discussed on graph privacy-preserving related work \cite{koutra2011algorithms}.
Such distance functions include the number of common neighbours of given nodes, cosine similarity, Jaccard similarity and so on. 
However, few of them are discussed in depth regarding adversarial behaviors (adversarial cost in game theory). 
In general, an attacker aims to make ``minimal" perturbations on the existing graph and therefore such distance measurement is important to measure the quality of attacks. How to design and choose proper distance function to quantify the attack ability under different attack scenarios is also critical towards developing defensive approaches regarding specific threat model. 
We will discuss potential perturbation evaluation metrics in detail in Sec~\ref{sec:advmetric}.

In addition to the unique properties of each graph distance function, it would also be interesting to analyze the ``equivalence" among them.
For instance, an attacker aims to attack one node by adding/removing one edge in the graph can encounter similar ``adversarial cost" as adding/removing edges.
It is not hard to see that by using a graph distance function or similarity measures,
only a few targets would be the optimal choices for the attacker ({\em with different distance}), so this can also help to optimize the adversarial targets.
In summary, due to the complexity and diversity of graph representations and adversarial behaviors,
perturbation evaluation or graph similarity measurement will depend on various factors such as different learning tasks, adversarial strategies, and adversarial cost types.






\subsection{Adversarial Perturbation}
\label{sec:advmetric}

To generate adversarial samples on graph data, we can modify the nodes or edges from the original graph.
However, the modified graph $\widehat{G}$ need to be ``similar'' with the original graph $G$ based on certain perturbation evaluation metrics and remain ``imperceptible". The following metrics help understand how to define ``imperceptible perturbation".

\textbf{Edge-level Perturbation.}
In most existing works, the attacker is capable of adding/removing/rewiring edges in the whole original graph within a given budget.
In this case, the number of modified edges is usually used to evaluate the magnitude of perturbation.
In addition to other perturbations, edge perturbation is hardly found by the defender, especially in dynamic graphs.

\textbf{Node-level Perturbation.}
The attacker is also capable of adding/removing nodes, or manipulating the features of target nodes.
The evaluation metric in this case can be calculated based on the number of nodes modified or the distance between the benign and adversarial feature vectors. 

\textbf{Structure Preserving Perturbation.}
Similar to edge-level perturbation, an attacker can modify edges in the graph within a given budget in terms of graph structure.
Compared to general edge-level perturbation, this considers more structural preservation, such as total degree, node distribution, etc.
For instance, in \cite{zugner2018adversarial}, the attacker is required to preserve the key structural features of a graph such as the degree distribution.
Therefore, the perturbation here can be measured by the graph structure drift.

\textbf{Attribute Preserving Perturbation.}
In the attributed graphs, each node or edge has its own features. 
In addition to manipulating the graph structure,
the attacker can choose to modify the features of nodes or edges to generate adversarial samples on graph data.
Various measurements based on graph-attribute properties can be analyzed to characterize the perturbation magnitude. 
For instance, in \cite{zugner2018adversarial}, 
the authors argue adding a feature is imperceptible if a probabilistic random walker on the co-occurrence graph can reach it with high probability by starting from existing features.

Note that, most GNN methods learn the feature representation of each node, which means it could be easily attacked by structure-only, feature-only perturbations or both. 

\textbf{Principles of imperceptible perturbation evaluation.}
Given various graph distance discussion, there is no clear discussion in existing research about how to set the adversarial cost for attacks on graph data so far. Therefore, we summarize some principles of defining the perturbation evaluation metrics as below for future research.
\begin{itemize}
    \item For static graph, both the number of modified edges and the distance between the benign and adversarial feature vectors should be small.
    \item For a dynamic graph, we can set the distance or adversarial cost based on the intrinsic changing information over time.
    For example, by using statistic analysis, we can get the upper bound of the information manipulated in practice,
    and use this information to set an imperceptible bound. 
    \item For various learning tasks on graph data, e.g., node or graph classification,
    we need to use a suitable graph distance function to calculate the similarity between the benign and its adversarial sample.
    For example, we can use the number of common neighbours to evaluate the similarity of two nodes, but this is not applicable for two individual graphs. 
\end{itemize}
In summary, compared to image and text data, an attacker first can modify more features on the information network, and also can explore more angles to define ``imperceptible'' based on the format of graph data and the application task.

\subsection{Attack Stage}

The adversarial attacks can happen at two stages:
evasion attack (model testing) and poisoning attacks (model training).
It depends on the attacker’s capacity to insert adversarial perturbations:

\textbf{Poisoning Attack.}
Poisoning attack tries to affect the performance of the model by adding adversarial samples into the training dataset.
Most existing works are poisoning attacks,
and their node classification tasks are performed in the transductive learning setting.
In this case, once the attacker changes the data, the model is retrained.
Mathematically, by setting $G'_j = \widehat{G}^{c_j}$ in Eq.~(\ref{eq:uniformattack}),
we have a general formula for adversarial attack on graph data under poisoning attacks.

\textbf{Evasion Attack.}
Evasion attack means that the parameters of the trained model are assumed to be fixed.
The attacker tries to generate the adversarial samples of the trained model.
Evasion attack only changes the testing data, which does not require to retrain the model.
Mathematically, by setting $G'_j$ to original $G_j$ in Eq.~(\ref{eq:uniformattack}),
we have a general formula for adversarial attack on graph data under evasion attacks.

\subsection{Attack Objective}

Though all adversarial attacks are modifying the data,
an attacker needs to choose their attack targets or objectives: model or data.
In this case, we can summarize them as model objective and data objective.

\textbf{Model Objective.}
Model objective is attacking a particular model by using various approaches.
It could be either evasion attack or poisoning attack.
Most current adversarial attack is related to model objective attack.
The target could be either GNN or other learning models. An attacker wants to make the model become non-functional in multiple scenarios.
Model objective attack can be categorized by whether using the gradient information of the model or not.
\begin{itemize}
    \item \textbf{Gradient-based Attack.} 
    In most studies, we can see that the gradient-based attack is always the simplest and most effective approach.
    Most gradient-based attack, no matter white-box or black-box, tries to get or estimate the gradient information to find the most important features to the model.
    Based on the above knowledge, an attacker can choose to modify the limited information based on the feature importance to the model and make the model inaccurate when using the modified information \cite{dai2018adversarial,zugner2018adversarial,bojchevski2019adversarial}.
    \item \textbf{Non-gradient-based Attack.} 
    In addition to gradient information, an attack could destroy the model without any gradient information.
    As we know, besides the gradients, many reinforcement learning based attack methods can attack the model based on long-term rewards \cite{sun2018data,dai2018adversarial,ma2019attacking}.
    Some works can also construct the adversarial samples with generative models \cite{bose2019generalizable,chen2019multiscale,fox2019robust}.
    All the above approaches can attack the model without the gradient information but attack the model in practice. 
\end{itemize}

\textbf{Data Objective.}
Unlike model objective attacks, data objective attacks do not attack a specific model.
Such attacks happen when the attacker only has access to the data, but does not have enough information about the model. 
In general there are two settings when data become the target.
\begin{itemize}
    \item \textbf{Model Poisoning.} Unsupervised feature analysis approaches can still get useful information from the data without any knowledge of the training approach.
    Even with a small perturbation on the data, it can make general training approaches cease to work. Besides, backdoor attack is another relevant hot topic where an attacker only injects the adversarial signals in the dataset, but does not destroy the model performance on regular samples~\cite{xi2020graph, zhang2020backdoor}.
    \item \textbf{Statistic Information.}
    In addition to using the data to train a model, in many studies, researchers use statistical results or simulation results from the graph data \cite{waniek2018hiding,zhou2019attacking,dey2019manipulating}. In this case, an attacker can break the model based on the capturing of the valuable statistical information on graph data. For example, by modifying a few edges between different communities based on structural information and analysis, one can make communities counting inaccurate under this attack \cite{waniek2018hiding}. 
\end{itemize}

\subsection{Attack Knowledge}

The attacker would receive different information to attack the system.
Based on this, we can characterize the dangerous levels of existing attacks.

\textbf{While-box Attack.}
In this case, an attacker can get all information and use it to attack the system,
such as the prediction result, gradient information, etc.
The attack may not work if the attacker does not fully break the system first.

\textbf{Grey-box Attack.} 
An attacker gets limited information to attack the system.
Comparing to white-box attack, it is more dangerous to the system, since the attacker only need partial information.

\textbf{Black-box Attack.}
Under this setting, an attacker can only do black-box queries on some of the samples.
Thus, the attacker generally can not do poisoning attack on the trained model.
However, if black-box attack can work, it would be the most dangerous attack compared with the other two,
because the attacker can attack the model with the most limited acknowledge.

Most existing papers only study white-box attack on the graph,
and there are lots of opportunities to study other attacks with different levels of knowledge.

\subsection{Attack Goal}

Generally, an attacker wants to destroy the performance of the whole system,
but sometimes they prefer to attack a few important target instances in the system.
Based on the goal of an attack, we have:

\textbf{Availability Attack.}
The adversarial goal of availability attack is to reduce the total performance of the system.
For example, by giving a modification budget,
we want the performance of the system decreasing the most as the optimal attack strategy.

\textbf{Integrity Attack.} 
The adversarial goal of integrity attack is to reduce the performance of target instances.
For example, in recommendation systems,
we want the model to not successfully predict the hidden relation between two target users.
However, the total performance of the system is the same or similar to the original system.

Availability attack is easier to detect than integrity attack under the positioning attack setting. Therefore, meaningful availability attack studies are in general under the evasion attack setting.

\subsection{Attack Task}

Corresponding to various tasks on graph data,
we show how to attack each task and explain the general idea by modifying the unified formulation. 

\textbf{Node-relevant Task.}
As mentioned before, most attack papers focus on node-level tasks,
including node classification \cite{dai2018adversarial, zugner2018adversarial,zugner2019adversarial,wang2019attacking,wu2019adversarial,chang2020restricted,xu2019topology} and node embedding \cite{bojchevski2019adversarial,zhang2019towards}.
The main difference is that node embedding uses the low dimensional representations of each node for an adversarial attack.
Mathematically, by setting $c_{i}$ as representation of node target in Eq.~(\ref{eq:uniformattack}),
we have a general formula for adversarial attack on node-relevant tasks.

\textbf{Link-relevant Task.}
Several other existing works study node embedding~\cite{bojchevski2019adversarial, chen2018link, sun2018data} or topological similarity~\cite{waniek2018attack, zhou2019attacking} and use them for link prediction.
Compared with node classification,
link prediction requires to use different input data,
where $c_{i}$ represents link target, i.e., the information of a pair of nodes.
By setting $c_{i}$ as representation of link target and $y_i \in [0, 1]$ in Eq.~(\ref{eq:uniformattack}),
we have a general formula for adversarial attack on link-relevant tasks.

\textbf{Graph-relevant Task.}
Compared with node classification, graph classification needs the graph representation instead of the node representation~\cite{dai2018adversarial, tang2020adversarial, zhang2020backdoor, xi2020graph}.
By setting $c_{i}$ as representation of graph target in Eq.~(\ref{eq:uniformattack}),
we have a general formula for adversarial attack on graph-relevant tasks.

\subsection{Summary: Attack on Graph}
In this subsection, we analyze the contributions and limitations of existing works.
Then we discuss the potential research opportunities in this area.

\textbf{Contributions.}
First, we list all released papers and their characteristics in Table \ref{table::attack_summary}, and then categorize them into selected main topics in Table \ref{table::attack_catagory}.
Then, we summarize the unique contributions of existing adversarial attacks.
Note that, because 11 of 34 papers we discuss are pre-print version, we especially list the venue in Table \ref{table::attack_summary}.
We also firstly use \textit{Strategy} and \textit{Approach} to differ individual attack method. \textit{Strategy} refers to the high-level design philosophy of an attack, while \textit{Approach} represents the concrete approach the attacker takes to perturb the graph data.

\textbf{Graph Neural Networks.}
Most adversarial attacks are relevant to graph neural networks.
\cite{dai2018adversarial} used reinforcement learning approach to discover adversarial attack,
which is the only approach that supports black-box attack compared to other works.
\cite{zugner2018adversarial} studied adversarial graph samples with traditional machine learning
and deep learning. Meanwhile, they are the first and only group to discuss the adversarial attack on attributed graph. \cite{chen2018link, sun2018data} mainly attacked the link predication task with a deep graph convolutional embedding model.
\cite{bojchevski2019adversarial} attacked multiple models by approximating the spectrum and using the gradient information.
\cite{wang2019attacking} attacked node classification through optimization approach and systematically discussed adversarial attacks on graph data.
Previous works focused on edge or node modification, whereas \cite{wu2019adversarial} also modified the node features and proposed a hybrid attack on the graph convolutional neural networks (GCN) \cite{kipf2016semi}.
In addition to gradient check, \cite{xu2019topology,entezari2020all} attacked GCN by using the first-gradient optimization and low-rank approximation which makes an attack more efficient.
\cite{chang2020restricted} attacked general learning approaches by devising new loss and approximating the spectrum. 
\cite{hou2019alphacyber} used graph attack knowledge into the malware detection problem, which showed various graph-based applications to be vulnerable to adversarial attacks.
Without gradient check and optimization design, \cite{sun2019node} used reinforcement learning to attack GCN. However, it contains an obvious issue that it needs to break the graph structure by injecting new nodes.
\cite{li2020adversarial} tried to hide nodes in the community by attacking the graph auto-encoder model.
Instead of using a gradient check or other optimization approaches, this work leverage the surrogate community detection model to achieve the attacking goal.
More recent works investigates the vulnerability of GNNs under backdoor attacks~\cite{xi2020graph, zhang2020backdoor}. Backdoor attack modifies the labels of the triggers (e.g., subgraphs with typical patterns) in the training data, and it aims to make the GNNs misclassify those triggers without affecting the overall performance of GNNs on the testing data. 

\textbf{Others.}
Though many attack works are relevant to GNN, many recent papers start to focus on other types of adversarial attacks on graph data.
\cite{chen2017practical} is one of the first works to attack the graph data, and it also first proposed the attack approach in the unsupervised learning setting.
\cite{waniek2018hiding} first attacked community detection though edge rewriting based on a heuristic approach.
\cite{waniek2018attack} attacked link prediction based on a heuristic approach which is based on the similarity measures.
\cite{zhou2019attacking} used a greedy approach to attack link prediction based local and global similarity measure.
In addition to traditional graph applications, \cite{zhang2019towards} first attacked knowledge graph and destroyed the basic relational graph prediction model.
\cite{chen2019ga} attacked community detection based on genetic algorithms. Unlike previous approaches, it chose to use rewiring instead of adding/removing edges while attacking the data.
\cite{fox2019robust} used a generation approach to create a new isomorphism network to attack node classification.
In addition to all previous works, \cite{dey2019manipulating} started to study attacks through theoretical analysis, and we believe more theoretical works will be seen in this domain.
They can help us understand the attacks better on graph data.
Besides the applications mentioned above, attacking graphs in recommender system~\cite{fang2018poisoning, peng2020robust, zhang2020gcn}, fraud detection~\cite{breuer2020friend, dou2020robust}, opinion dynamic~\cite{gaitonde2020adversarial, chen2020network}, and graph classification~\cite{tang2020adversarial, zhang2020backdoor, xi2020graph} tasks have been drawing attention from researchers as well.

\textbf{Limitations.}
The limitations of most current works are summarized below. 
Most existing works do not give very clear strategies about the setting of the budget and distance with reasonable explanations in real applications.
Different from other adversarial attacks, most graph modifications can hardly be noticed by humans in real life.
To solve this problem, we give a more detailed discussion on perturbation and evaluation metrics in Section \ref{sec:metric}.
Meanwhile, about graph imperceptible evaluation metrics, 
most papers \cite{bojchevski2019adversarial, chen2018link, dai2018adversarial} use one metric for attack,
but these adversarial samples could be detected by other existing imperceptible evaluation metrics.
In this work, we list all existing evaluation metrics,
and recommend future adversarial samples to be imperceptible with more listed evaluation metrics.
Another main issue is due to the different problem formulations.
To this end, we give the unified problem formulation for all existing works discussed in this survey.

\textbf{Most Recent Work.} For the recently proposed attack methods, imperciptible pertubations such as added edges and modified node features are also the principle approaches like previous work. For the defense models, instead of commonly-used adversarial training techniques before, some researchers \mbox{\cite{liu2021graph, liu2021elastic, chen2021understanding}} first tried to propose new neiborhood aggregation schemas which guarantee the theoretical robustness under adversarial attack.

\textbf{Future Directions.}
Adversarial attack on graph data is a new and hot area, and potential research opportunities are summarized below:
1) Most graphs are associated with attributes or more complex contents on nodes or edges in practice. However, few studies have well designed adversarial attack on attributed graphs, e.g., heterogeneous information networks and web graphs.
2) Some advanced ideas can be applied for generating the adversarial samples, e.g., homomorphism graph.
3) Various learning settings are not sufficiently studied yet, such as graph-level attacks and inductive learning on node-level attacks.
4) Most existing attacks do not consider various imperceptibility metrics in their models.
Concise and comprehensive imperceptibility metrics are necessary in different tasks.
A good and explainable evaluation metric may easily discover more existing adversarial samples created by current methods. 
5) Last but not least, the distance or similarity measures of high quality adversarial samples are not well studied in this area.





\begin{table*}[hbt!]
\centering
\caption{Summary of adversarial defense works on graph data (time ascending).}
\label{table::defense_summary}
\resizebox{\linewidth}{!}{  
\begin{tabular}{cccccccccc}
\toprule
Task &Ref. & Year & Venue &  Model  &  Corresp. Attack  &  Strategy  & Baseline & Metric & Dataset\\ 
\hline


\multirow{61}{*}{Node classification}

&\cite{feng2019graph} & 2019 & TKDE &  GCN  & - & \begin{tabular}[c]{@{}c@{}}Adversarial\\ training  \end{tabular}& \begin{tabular}[c]{@{}c@{}c@{}c@{}}DeepWalk, GCN, \\ Planetoid, LP, \\ GraphVAT,\\ GraphSCAN\end{tabular} & Accuracy &  \begin{tabular}[c]{@{}c@{}c@{}}Cora, \\NELL, \\ Citeseer\end{tabular} \\ 
\cline{2-10}

&\cite{dai2019adversarial} & 2019 & WWW &  DeepWalk  &  - & \begin{tabular}[c]{@{}c@{}}Adversarial \\ training\end{tabular}  & \begin{tabular}[c]{@{}c@{}c@{}}DeepWalk, LINE\\ Node2vec, GraRep,\\Graph Factorization\end{tabular}  & \begin{tabular}[c]{@{}c@{}}Accuracy, \\ AUC\end{tabular} & \begin{tabular}[c]{@{}c@{}c@{}c@{}}Cora, Wiki\\ Citeseer,\\CA-GrQc, \\CA-HepTh\end{tabular} \\
\cline{2-10}

&\cite{wang2019graphdefense} & 2019 & arXiv  &   \begin{tabular}[c]{@{}c@{}}GCN,\\ GraphSAGE\end{tabular}  &   -  &  \begin{tabular}[c]{@{}c@{}}Adversarial\\ training \end{tabular}  &  \begin{tabular}[c]{@{}c@{}c@{}}Drop edges, Discrete\\adversarial\\ training \end{tabular} &  \begin{tabular}[c]{@{}c@{}c@{}}Accuracy, \\ Correct \\classification rate \end{tabular} & \begin{tabular}[c]{@{}c@{}c@{}}Cora, \\Citeseer,\\Reddit\end{tabular}\\
\cline{2-10}

&\cite{jin2019latent} & 2019 & \begin{tabular}[c]{@{}c@{}}ICML\\ Workshop\end{tabular}  &  GCN   &  Nettack   &  \begin{tabular}[c]{@{}c@{}} Adversarial\\ training \end{tabular} &  \begin{tabular}[c]{@{}c@{}}GCN, SGCN,\\FastGCN, SGC \end{tabular}  & \begin{tabular}[c]{@{}c@{}}ASR,\\ Accuracy \end{tabular}  &  \begin{tabular}[c]{@{}c@{}c@{}}Citeseer, Cora, \\ Pubmed, Cora-ML,\\ DBLP, PolBlogs\end{tabular}\\ 
\cline{2-10}

&\cite{deng2019batch} & 2019 &  \begin{tabular}[c]{@{}c@{}}ICML\\ Workshop\end{tabular}  & GCN & - & \begin{tabular}[c]{@{}c@{}} Adversarial\\ training \end{tabular}	 & \begin{tabular}[c]{@{}c@{}c@{}}GCN, GAT, LP,\\ DeepWalk, Planetoid,\\ Monet, GPNN \end{tabular}  & Accuracy & \begin{tabular}[c]{@{}c@{}}Citeseer, Cora, \\ Pubmed, NELL\end{tabular}\\ 
\cline{2-10}

&\cite{sun2019virtual} & 2019 & PRCV &  GCN  & -  & \begin{tabular}[c]{@{}c@{}c@{}}Virtual \\adversarial \\ training\end{tabular}  & GCN  & Accuracy & \begin{tabular}[c]{@{}c@{}c@{}}Cora,\\ Citeseer,\\Pubmed\end{tabular} \\
\cline{2-10}

&\cite{bojchevski2019certifiable} & 2019 & NIPS  &  GCN   &  -    &  \begin{tabular}[c]{@{}c@{}c@{}} Robust training,\\ MDP to\\ get bound\end{tabular} &    GNN &\begin{tabular}[c]{@{}c@{}c@{}}Accuracy, \\Worst-case \\margin\end{tabular} &\begin{tabular}[c]{@{}c@{}c@{}}Cora-ML,\\ Pubmed,\\Citeseer\end{tabular}\\ 
\cline{2-10}

&\cite{xu2019topology} & 2019 & IJCAI  &   GCN  &  \begin{tabular}[c]{@{}c@{}}DICE,\\ Meta-self\end{tabular}    &  \begin{tabular}[c]{@{}c@{}c@{}}Check gradients,\\Adversarial\\ training\end{tabular}   &  GCN  &  \begin{tabular}[c]{@{}c@{}c@{}}Accuracy,\\Misclassification\\ rate\end{tabular}  & \begin{tabular}[c]{@{}c@{}}Cora, \\Citeseer\end{tabular}\\ 
\cline{2-10}




&\cite{wu2019adversarial} & 2019 & IJCAI &   GCN  &   \begin{tabular}[c]{@{}c@{}c@{}}Random,\\ Nettack \\ FGSM, JSMA\end{tabular}  &  Drop edges   &  GCN  & \begin{tabular}[c]{@{}c@{}c@{}}Accuracy, \\ Classification \\margin\end{tabular}   & \begin{tabular}[c]{@{}c@{}c@{}}Cora,\\ Citeseer,\\PolBlogs\end{tabular}\\ 
\cline{2-10}

&\cite{zugner2019certifiable} & 2019 & KDD  &  GCN, GNN   &  -   &  \begin{tabular}[c]{@{}c@{}} Convex\\ optimization \end{tabular} &  GNN  &  \begin{tabular}[c]{@{}c@{}c@{}}Accuracy, \\Average\\ worst-case margin\end{tabular} &\begin{tabular}[c]{@{}c@{}c@{}}Cora-ML,\\ Pubmed,\\Citeseer\end{tabular} \\ 
\cline{2-10}

&\cite{miller2019improving} & 2019 & \begin{tabular}[c]{@{}c@{}}KDD\\Workshop\end{tabular} &   \begin{tabular}[c]{@{}c@{}}GCN,\\ Node2vec\end{tabular}  &  -   &  \begin{tabular}[c]{@{}c@{}}Change \\training set  \end{tabular} &  \begin{tabular}[c]{@{}c@{}}GCN,\\ Node2vec\end{tabular}   & \begin{tabular}[c]{@{}c@{}c@{}}Adversary budget, \\Classification \\ margin \end{tabular} & Cora, Citeseer\\
\cline{2-10}

&\cite{zhu2019robust} & 2019 & KDD &  GCN   &   \begin{tabular}[c]{@{}c@{}c@{}}Nettack, \\RL-S2V,\\ Random\end{tabular}   &  \begin{tabular}[c]{@{}c@{}c@{}c@{}}Gaussian \\distribution layer,\\ Variance-based\\ attention\end{tabular}		  & GCN, GAT  & Accuracy  & \begin{tabular}[c]{@{}c@{}c@{}}Cora,\\ Citeseer,\\Pubmed\end{tabular}\\
\cline{2-10}

&\cite{tang2020transferring} & 2020 & WSDM  &  GNN   &   Metattack  &   \begin{tabular}[c]{@{}c@{}c@{}}Meta learning, \\Transfer  from \\clean graph\end{tabular}  &  \begin{tabular}[c]{@{}c@{}}GCN, GAT, \\ RGCN, VPN\end{tabular}   &  Accuracy  & \begin{tabular}[c]{@{}c@{}c@{}}Pubmed, \\Yelp,\\Reddit\end{tabular} \\ 
\cline{2-10}

&\cite{entezari2020all} & 2020 & WSDM &  GCN, t-PINE   &  \begin{tabular}[c]{@{}c@{}}Nettack,\\ LowBlow	\end{tabular} &    \begin{tabular}[c]{@{}c@{}}Low-rank \\approximation	\end{tabular}	  &   - &  \begin{tabular}[c]{@{}c@{}}Correct\\ classification rate   \end{tabular}& \begin{tabular}[c]{@{}c@{}} Cora-ML, Citeseer, \\PolBlogs \end{tabular}\\  \cline{2-10}

&\cite{jin2020graph} & 2020 & KDD &  GNN  &  \begin{tabular}[c]{@{}c@{}c@{}}Nettack, \\ Meta-self,\\ Random\end{tabular}	  &  \begin{tabular}[c]{@{}c@{}} Graph structure \\learning\end{tabular}   &   \begin{tabular}[c]{@{}c@{}c@{}}GCN, GCN-SVD,\\ RGCN, GAT, \\ GCN-Jaccard \end{tabular}   &  Accuracy  & \begin{tabular}[c]{@{}c@{}c@{}}Cora, Pubmed,\\ Polblogs,\\Citeseer \end{tabular}\\ \cline{2-10}

&\cite{chang2021not} & 2021 & NIPS & GCN & \begin{tabular}[c]{@{}c@{}}Nettack-One, \\ Nettack-Multi, \\Metattack \end{tabular} &\begin{tabular}[c]{@{}c@{}}Transfer robustness of \\ low-frequency \\ components by co-training\end{tabular} &\begin{tabular}[c]{@{}c@{}}GCN-Jaccard/-SVD \\ GNN GUARD, \\Pro-GNN \end{tabular} &Accuracy & \begin{tabular}[c]{@{}c@{}}Cora, Citeseer, \\ Pubmed, Coauthor CS, \\Amazon Photo \end{tabular}\\ \cline{2-10}

&\cite{liu2021graph} & 2021 & NIPS & GNN & Nettack &\begin{tabular}[c]{@{}c@{}}Adaptive message passing \\ against abnormal \\node features \end{tabular} &\begin{tabular}[c]{@{}c@{}}GCN, GAT \\ APPNP, \\GCNII \end{tabular} & Accuracy & \begin{tabular}[c]{@{}c@{}}Cora, Citeseer, Pubmed,\\ Coauthor CS/Physics, \\Amazon Computers/Photo \end{tabular}\\ \cline{2-10}

&\cite{chen2021understanding} &2021 &IJCAI &GCN &  Nettack &\begin{tabular}[c]{@{}c@{}} Aggregation with a \\ high breakdown point \end{tabular} & \begin{tabular}[c]{@{}c@{}}GCN, RGCN,\\ GCN-Jaccard, SimPGCN \end{tabular} &Accuracy &\begin{tabular}[c]{@{}c@{}}Cora, Cora-ML,\\ Citeseer, Pubmed \end{tabular} \\ \cline{2-10}

&\cite{jin2021node} &2021 &WSDM &GNN &Metattack 
&\begin{tabular}[c]{@{}c@{}}Information aggregation \\based on feature similarity \end{tabular} &\begin{tabular}[c]{@{}c@{}} GCN, GAT,\\Pro-GNN, GCN-Jaccard \end{tabular} &Accuracy &\begin{tabular}[c]{@{}c@{}}Cora, Citeseer, \\Pubmed \end{tabular}\\ \cline{2-10}

&\cite{liu2021elastic} &2021 &ICML &GNN &Metattack &\begin{tabular}[c]{@{}c@{}}Message passing \\ with graph smoothing \end{tabular} &\begin{tabular}[c]{@{}c@{}}GCN, GAT, ChebNet, \\ GraphSAGE, APPNP, SGC \end{tabular} &Accuracy & \begin{tabular}[c]{@{}c@{}}Cora, Citeseer, Pubmed,\\ Coauthor CS/Physics, \\Amazon Computers/Photo \end{tabular}\\ \cline{2-10}

&\cite{zhang2021detection} &2021 &AISTATS &GCN 
& \begin{tabular}[c]{@{}c@{}}DICE, Nettack, \\GF-Attack \end{tabular}
& Detect malicious nodes &\begin{tabular}[c]{@{}c@{}}GCN,SGCN,GAT, \\RGCN, GCN-Jaccard/-SVD \end{tabular} & Accuracy, AUC & \begin{tabular}[c]{@{}c@{}}Cora, Citeseer, \\Polblogs, Pubmed \end{tabular}\\ \cline{2-10}

&\cite{xu2021speedup} &2021 &CIKM & GCN &\begin{tabular}[c]{@{}c@{}}Metattack, \\ Nettack, \\Random \end{tabular} & \begin{tabular}[c]{@{}c@{}}Graph structure \\learning with low-rank \\prior knowledge \end{tabular} & \begin{tabular}[c]{@{}c@{}}RGCN, Pro-GNN \\ GCN-Jaccard/-SVD \end{tabular} & Accuracy & \begin{tabular}[c]{@{}c@{}}Cora, Citeseer \\ Polblogs, Pubmed \end{tabular} \\
\hline

\multirow{7}{*}{Link prediction}

&\cite{pezeshkpour2019investigating} & 2019 & NAACL &  \begin{tabular}[c]{@{}c@{}c@{}}Knowledge \\graph \\ embeddings\end{tabular}   &  -   &  \begin{tabular}[c]{@{}c@{}}Adversarial \\modification \end{tabular}  &  -  &  Hits$@$K, MRR  & \begin{tabular}[c]{@{}c@{}c@{}}Nations, WN18, \\Kinship, \\YAGO3-10\end{tabular}\\ \cline{2-10}

&\cite{yu2019target} & 2019 & TKDE  &   \begin{tabular}[c]{@{}c@{}c@{}}Link \\ prediction\\ methods\end{tabular}  &  \begin{tabular}[c]{@{}c@{}c@{}}Resource \\ Allocation\\ Index\end{tabular}    &  \begin{tabular}[c]{@{}c@{}c@{}}Estimation of  \\Distribution\\ Algorithm\end{tabular}   &  \begin{tabular}[c]{@{}c@{}}RLR, RLS \\HP, GA\end{tabular}  &  \begin{tabular}[c]{@{}c@{}}Precision,\\AUC\end{tabular}  & \begin{tabular}[c]{@{}c@{}c@{}}Mexican, Dolphin\\Bomb, Lesmis, \\ Throne, Jazz\end{tabular}\\ \cline{2-10}

&\cite{zhou2019adversarial} & 2019 & ICDM  & \begin{tabular}[c]{@{}c@{}}Similarity \\measures\end{tabular}    &   -  &  \begin{tabular}[c]{@{}c@{}c@{}}Bayesian \\Stackelberg game \\and optimization\end{tabular}   &  \begin{tabular}[c]{@{}c@{}c@{}}Protect \\ Potential\\ Neighbors\end{tabular}  &  \begin{tabular}[c]{@{}c@{}}Damage \\ prevention ratio\end{tabular}   &\begin{tabular}[c]{@{}c@{}c@{}}PA, \\TV Show,\\PLD, Gov \end{tabular} \\ 
\hline

Graph classification &\cite{zhang2020backdoor}& 2020 & arXiv & GIN  & Graph generation  &    \begin{tabular}[c]{@{}c@{}}Randomized\\ subsampling\end{tabular}  &  -  &  \begin{tabular}[c]{@{}c@{}c@{}}ASR,\\Clean accuracy, \\Backdoor accuracy\end{tabular}  & \begin{tabular}[c]{@{}c@{}c@{}}Twitter,\\Bitcoin, \\COLLAB\end{tabular}\\ \hline

Node embedding &\cite{chen2019can} & 2019 & arXiv &  GNN  & \begin{tabular}[c]{@{}c@{}}Nettack, \\ FGA\end{tabular}  & \begin{tabular}[c]{@{}c@{}}Smoothing\\ gradients\end{tabular} & \begin{tabular}[c]{@{}c@{}}Adversarial \\training\end{tabular}  & ADR, ACD & \begin{tabular}[c]{@{}c@{}}Cora, Citeseer,\\PolBlogs\end{tabular} \\ \hline

\begin{tabular}[c]{@{}c@{}}Malware detection, \\ Node classification \end{tabular} &\cite{hou2019alphacyber} & 2019 & CIKM &  \begin{tabular}[c]{@{}c@{}c@{}}Heterogeneous \\graph, \\
Metapath2vec\end{tabular}  &  -   &  \begin{tabular}[c]{@{}c@{}}Attention\\ mechanism \end{tabular}  &  \begin{tabular}[c]{@{}c@{}}Other malware \\ detection algs\end{tabular}  & \begin{tabular}[c]{@{}c@{}}Accuracy, F1, \\ Precision, Recall \end{tabular} & \begin{tabular}[c]{@{}c@{}}Private\\ dataset \end{tabular}\\ 
\hline


Community detection &\cite{jia2020certified} & 2020 & WWW  &  \begin{tabular}[c]{@{}c@{}}Community\\ detection algs\end{tabular}  &  -	  &  \begin{tabular}[c]{@{}c@{}c@{}} Robust \\certification \\ with optimization\end{tabular}   &  -  &  \begin{tabular}[c]{@{}c@{}}Certified\\ accuracy\end{tabular}  & \begin{tabular}[c]{@{}c@{}}Email, DBLP,\\ Amazon\end{tabular}\\ 
\hline

\multirow{4}{*}{Fraud detection}
&\cite{breuer2020friend} & 2020 & WWW  &  \begin{tabular}[c]{@{}c@{}}Graph-based\\Sybil detectors \end{tabular} &  \begin{tabular}[c]{@{}c@{}}Change label, \\ Graph generation\end{tabular} &  \begin{tabular}[c]{@{}c@{}} Probability \\ estimation\end{tabular}   &   \begin{tabular}[c]{@{}c@{}c@{}}VoteTrust,\\ SybilRank, SybilSCAR, \\ SybilBelief \end{tabular}   &  AUC  & \begin{tabular}[c]{@{}c@{}c@{}}Facebook,\\Synthetic graphs \end{tabular}\\ \cline{2-10}

&\cite{dou2020robust} & 2020 & KDD &  \begin{tabular}[c]{@{}c@{}}Graph-based\\ Fraud detectors\end{tabular}  &  \begin{tabular}[c]{@{}c@{}c@{}}IncBP, IncDS,\\ IncPR, Random,\\Singleton\end{tabular}	  &  \begin{tabular}[c]{@{}c@{}c@{}} Minimax game, \\Reinforcement \\ learning\end{tabular}   &   \begin{tabular}[c]{@{}c@{}}SpEagle, GANG\\ Fraudar, fBox\end{tabular}   &  \begin{tabular}[c]{@{}c@{}}
Practical\\ effect\end{tabular}  & \begin{tabular}[c]{@{}c@{}c@{}}YelpChi, \\ YelpNYC,\\ YelpZip\end{tabular}\\ 
\hline

\begin{tabular}[c]{@{}c@{}} Manipulating \\ opinion \end{tabular} &\cite{gaitonde2020adversarial}& 2020 & arXiv  &  Graph model & -  &  \begin{tabular}[c]{@{}c@{}} Minimax game,\\ Convex optimization \end{tabular}  &  -  &  -  & -\\
\hline

\multirow{4}{*}{Recommendation system}

&\cite{zhang2020gcn} & 2020 & SIGIR  &  GCN  &  \begin{tabular}[c]{@{}c@{}c@{}}Mixed, Hate, \\ Average,\\ Random\end{tabular}	  &  Fraud detection   &   \begin{tabular}[c]{@{}c@{}c@{}}RCF, GCMC,\\ GraphRec, MF, \\ AutoRec, PMF \end{tabular}   &  RMSE, MAE  & \begin{tabular}[c]{@{}c@{}}Yelp, \\Moive\&TV \end{tabular}\\ \cline{2-10}

&\cite{zhang2021graph} &2021 &WWW
& GCN & \begin{tabular}[c]{@{}c@{}}Attribute Inference \\attack \end{tabular} & Differential privacy &\begin{tabular}[c]{@{}c@{}}BPR, GCN, \\Blurm, DPAE, DPNE, \\DPMF, RAP \end{tabular} &\begin{tabular}[c]{@{}c@{}}F1, NDCG@K, \\Hits@K \end{tabular} &ML-100K\\ \hline

\begin{tabular}[c]{@{}c@{}}Node classification, \\ Link prediction, \\Community detection \end{tabular} &\cite{xu2020unsupervised} & 2022 & AAAI  &\begin{tabular}[c]{@{}c@{}}DeepWalk, \\ GAE, \\ DGI \end{tabular} & - &\begin{tabular}[c]{@{}c@{}} Graph representation \\ learning with \\ optimization \end{tabular} & \begin{tabular}[c]{@{}c@{}}Dwns\_AdvT, RSC, \\ DGI-EdgeDrop, \\-SVD, -Jaccard \end{tabular} & \begin{tabular}[c]{@{}c@{}}Accuracy, \\ AUC, NMI \end{tabular} & \begin{tabular}[c]{@{}c@{}}Cora, \\Citeseer,\\ Polblogs \end{tabular} \\ \hline

\begin{tabular}[c]{@{}c@{}} Node classification,\\ Graph classification \end{tabular} &\cite{wang2021certified} &2021 &KDD &GNN &  Random &randomized smoothing & GCN, GAT & Certified accuracy &\begin{tabular}[c]{@{}c@{}}Cora, Citeseer, Pubmed,\\ MUTAG, Proteins, IMDB \end{tabular} \\
\hline

\begin{tabular}[c]{@{}c@{}}Graph classification \\ Graph matching \end{tabular} &\cite{zhao2021expressive} &2021 &ICML &\begin{tabular}[c]{@{}c@{}}Graph classification\\/matching models\end{tabular} &\begin{tabular}[c]{@{}c@{}}Radnom, NEA, \\ GMA, RL-S2V \end{tabular} & Constrain the norm of gradient & \begin{tabular}[c]{@{}c@{}}PAN, Pro-GNN, GRAND, \\ GCN-SVD, RoboGraph, GraphCL, \\GroupSort, BCOP, FINAL \\ REGAL, MOANA, DGMC,\\ CONE-Align, G-CREWE \end{tabular} & Accuracy & \begin{tabular}[c]{@{}c@{}}AS, CAIDA, DBLP, \\ BZR, BZR\_MD, MUTAG \end{tabular}\\
\hline

Graph matching &\cite{ren2021integrated} &2021 &ICML &\begin{tabular}[c]{@{}c@{}}Graph matching \\ algo\end{tabular}
&\begin{tabular}[c]{@{}c@{}}Random, NEA, \\ GMA\end{tabular}
&\begin{tabular}[c]{@{}c@{}}Maximize distances between \\matched nodes; separate \\intra-graph nodes\end{tabular}  &\begin{tabular}[c]{@{}c@{}}FINAL, REGAL,\\ MOANA, DGMC,\\ CONE-Align, G-CREWE \end{tabular} &Hits@K & \begin{tabular}[c]{@{}c@{}}AS, CAIDA, DBLP \end{tabular} \\
\hline

Network alignment &\cite{zhou2021robust} &2021 &WWW &\begin{tabular}[c]{@{}c@{}}Network alignment \\ algo \end{tabular} &\begin{tabular}[c]{@{}c@{}}Random, Meta-self, \\ GF, CD, GMA, \\LowBlow \end{tabular} & \begin{tabular}[c]{@{}c@{}}Neutralize adversarial nodes \\ to adversarial-free \end{tabular} & \begin{tabular}[c]{@{}c@{}}GCN-Jaccard, \\ GCN-SVD, \\Pro-GNN \end{tabular} &Precision & \begin{tabular}[c]{@{}c@{}}AS, SNS,  \\ DBLP \end{tabular} \\
\hline

\begin{tabular}[c]{@{}c@{}}Recommendation system, \\ Knowledge graph, \\Quantum chemistry \end{tabular} &\cite{liao2021information} &2021 &ICML & GNN & \begin{tabular}[c]{@{}c@{}}Neighborhood  \\ attack \end{tabular} &Adversarial training &ChebNet, GraphSAGE & F1, AUC, RMSE & \begin{tabular}[c]{@{}c@{}}Movielens-1M, FB15k-237, WN18RR, \\ Citeseer, Pubmed, QM9 \end{tabular}\\

\bottomrule
\end{tabular}
}
\end{table*}

\section{Adversarial Defense on Graph Data} 
\label{sec:defense}
With graph data, recent intensive studies on adversarial attacks have also triggered the research on adversarial defenses. Here we survey existing works in this line and classify them into the two popular categories of \textit{Adversarial Training} and \textit{Attack Detection}. 
After them, we use an additional \textit{Other Methods} subsection to summarize the remaining methods that do not fit into the two generic categories.

\subsection{Adversarial Training}
\label{sec:AD_training}
While adversarial training has been widely used by attackers to perform effective adversarial intrusion, the same sword can be used by defenders to improve the robustness of their models against adversarial attacks \cite{goodfellow2014explaining}. In the graph setting, we formulate the objective of adversarial defense by slightly modifying our unified formulation of adversarial attacks, \textit{i.e.}, Eq.~(\ref{eq:uniformattack}), as follows
\begin{equation}\label{eq:uniformdefense}
    \begin{aligned}
    & \underset{\theta}{\text{min}} \underset{\widehat{G}^{c_i} \in \Phi(G_i)}{\text{max}}
    & & \sum_i \cL(f_{\theta}(c_{i}, \widehat{G}^{c_i}), y_i)).
    \end{aligned}
\end{equation}
where meanings of the notations remain the same as defined in Section \ref{sec:attack}. 
The idea is to alternatively optimize two competing modules during training, where the attacker tries to maximize task-oriented loss by generating adversarial perturbations $\widehat{G}$ on the graph, and the defender tries to minimize the same loss by learning the more robust graph model parameters $\theta$ under the generated adversarial perturbations. In this way, the learned graph model is expected to be resistant to future adversarial attacks.

 \textbf{Structure Perturbations.}
The earliest and most primitive way of perturbing the graph is to randomly drop edges \cite{dai2018adversarial}. The joint training of such cheap adversarial perturbations is shown to slightly improve the robustness of standard GNN models towards both graph and node classification tasks.
One step further, \cite{xu2019topology} proposed a topology attack generation method based on projected gradient descent to optimize edge perturbation. The topology attack is shown to improve the robustness of the adversarially trained GNN models against different gradient-based attacks and greedy attacks \cite{wang2018attack, zhou2019attacking, xu2019topology} without sacrificing node classification accuracy on the original graph.
In the meantime, \cite{dai2019adversarial} proposed to learn the perturbations in an unsupervised fashion by maximizing the influence of random noises in the embedding space, which improved the generalization performance of DeepWalk \cite{perozzi2014deepwalk} on node classification.
Towards similarity-based link prediction, \cite{zhou2019adversarial} formalized a Bayesian Stackelberg game to optimize the most robust links to preserve with an adversary deleting the remaining links.

 \textbf{Attribute Perturbations.}
Besides links, \cite{feng2019graph, sun2019virtual, deng2019batch} also perturb node features to enable virtual adversarial training \cite{miyato2018virtual} that enforces the smoothness between original nodes and adversarial nodes. In particular, \cite{feng2019graph} designed a dynamic regularizer forcing GNN models to learn to prevent the propagation of perturbations on graphs, whereas \cite{sun2019virtual} smoothed GCN in its most sensitive directions to improve generalization. \cite{deng2019batch} further conducted virtual adversarial training in batch to perceive the connectivity patterns between nodes in each sampled subsets. 
\cite{wang2019adversarial} leveraged adversarial contrastive learning \cite{bose2018adversarial} to tackle the vulnerabilities of GNN models to adversarial attacks due to training data scarcity and applied conditional GAN to utilize graph-level auxiliary information.  
Instead of approximating the discrete graph space, \cite{wang2019graphdefense} proposed to directly perturb the adjacency matrix and feature matrix by ignoring the discreteness, whereas \cite{jin2019latent} proposed to focus on the first hidden layer of GNN models to continuously perturb the adjacency matrix and feature matrix. These frameworks are all shown to improve GNN models on the node classification task.

 \textbf{Attack-oriented Perturbation}
Based on existing network adversarial attack methods of FGA \cite{chen2018fast} and Nettack \cite{zugner2018adversarial}, \cite{chen2019can} designed the adversarial training pipelines with additional smooth defense strategies. The pipeline is shown to improve GNN models against different adversarial attacks on node classification and community detection tasks.
\cite{dou2020robust} employed reinforcement learning to train a robust detector against mixed attacks proposed in the paper.

\subsection{Attack Detection}
Instead of generating adversarial attacks during training, another effective way of defense is to detect and remove (or reduce the effect of) attacks, under the assumption that data have already been polluted. Due to the complexity of graph data, the connection structures and auxiliary features can be leveraged based on various ad hoc yet intuitive principles to essentially differentiate clean data from poison ones and combat certain types of attacks.

 \textbf{Graph Preprocessing.}
\cite{xu2018characterizing} proposed different approaches to detect potential malicious edges based on graph generation models, link prediction and outlier detection.
Instead of edges, \cite{ioannidis2019graphsac} proposed to filter out node sets contaminated by anomalous nodes based on graph-aware criteria computed on randomly drawn subsets of nodes; \cite{zhang2019comparing} proposed to detect nodes subject to topological perturbations (particularly by Nettack \cite{zugner2018adversarial}) based on empirical analysis on the discrepancy between the proximity distributions of nodes and their neighbors.
These models only rely on network topology for attack detection.
On attributed graphs, based on the observations that attackers prefer adding edges over removing edges and the edges are often added between dissimilar nodes, \cite{wu2019adversarial} proposed to compute the Jaccard Similarity to remove suspicious edges between suspicious nodes.
\cite{xu2020edog} sampled sub-graphs from the poisoned training data and then employed outlier detection methods to detect and filter adversarial edges.
All of these models can be used for graph preprocessing before training normal graph models like GNNs.

 \textbf{Model Training.}
Rather than direct detection of suspicious nodes or edges before training, several works designed specific attention mechanisms to dynamically uncover and down-weigh suspicious data during training. 
\cite{zhu2019robust} assumed high prediction uncertainty for adversarial nodes and computed the attention weights based on the embedding variance in a Gaussian-based GCN. 
\cite{tang2020transferring} suggested to train an attack-aware GCN based on ground-truth poisoned links generated by Nettack \cite{zugner2018adversarial} and transfer the ability to assign small attention weights to poisoned links based on meta-learning.

 \textbf{Robustness Certification.}
On the contrary of detecting attacks, \cite{bojchevski2019certifiable, zugner2019certifiable} designed robustness certificates to measure the safety of individual nodes under adversarial perturbation. In particular, \cite{bojchevski2019certifiable} considered structural perturbation while \cite{zugner2019certifiable} considered attribute perturbation. Training GNN models jointly with these certificates can lead to a rigorous safety guarantee of more nodes. From a different perspective, \cite{jia2020certified} derived the robustness certificate of community detection methods under structural perturbation.
\cite{kenlay2020stability} proved polynomial
spectral graph filters are stable under structural perturbation.

 \textbf{Complex Graphs}
Beyond traditional homogeneous graphs, \cite{pezeshkpour2019investigating} studied the sensitivity of knowledge graph link prediction models towards adversarial facts (links) and the identification of facts. 
\cite{hou2019alphacyber} studied the detection of poisoning nodes in heterogeneous graphs to enhance the robustness of Android malware detection systems.

\subsection{Other Methods}
Now we summarize the remaining graph adversarial defense algorithms that are neither based on adversarial training nor aiming at attack detection. We further group them into three subcategories based on their modifications to the graph data and graph models.

 \textbf{Data Modifications.}
We have presented several attack detection algorithms that can be used for modifying graph data, \textit{i.e.}, graph preprocessing \cite{xu2018characterizing, ioannidis2019graphsac, zhang2019comparing}. There exist methods that modify graph data without directly detecting attacks. 
Based on the insight that Nettack \cite{zugner2018adversarial} only affects the high-rank singular components of the graph, \cite{entezari2020all} proposed to reduce the effect of attacks by computing the low-rank approximation of the graphs before training GNN models.
\cite{fox2019robust} proposed an augmented training procedure by generating more structurally noisy graphs to train GNN models for improved robustness, and showed it to be effective for structural role identification of nodes. 
\cite{miller2020topological} analyzed the topological characteristics of graphs and proposed two training data selection techniques to raise the difficulty of effective adversarial perturbations towards node classification.
These methods are all based on graph topology alone, and they only modify the graph data instead of the graph models.
\cite{zhang2020defensevgae} leveraged variational graph autoencoders to reconstruct graph structures from perturbed graphs where the reconstructed graphs can reduce the effects of adversarial perturbations.

 \textbf{Model Modifications.}
On the contrary, there exist methods that only modify the graph models, such as model-structure redesign or loss-function redesign.
The simplest way is to redesign the loss function.
From several existing works, the results show some loss functions perform better performance against the adversarial examples.
For example, \cite{jin2019power} designed an alternative operator based on graph powering to replace the classical Laplacian in GNN models with improved spectral robustness. They demonstrated the combination of this operator with vanilla GCN to be effective in node classification and defense against evasion attacks.
\cite{peng2020robust} proposed a hierarchical GCN model to aggregate neighbors from different orders and randomly dropped neighbor messages during the aggregation.
Such mechanism could improve the robustness of GCN-based collaborative filtering models.
\cite{zhang2020gnnguard} introduced neighbor importance estimation and the layer-wise graph memory components which can be integrated with GNNs.
Those two components could help increase the robustness of GNN models agaisnt various attacks.

\textbf{Hybrid Modifications.}
One step further, some methods modify both the graph data and graph models.
\cite{ioannidis2019edge} designed an edge-dithering approach to restoring unperturbed node neighborhoods with multiple randomly edge-flipped graphs and proposed an adaptive GCN model that learns to combine the multiple graphs. The proposed framework is shown to improve the performance and robustness of GCN towards node classification (in particular, protein function prediction) on attributed graphs.
\cite{miller2019improving} proposed a heuristic method to iteratively select training data based on the degrees and connection patterns of nodes. It further proposed to combine node attributes and structural features and use SVM for node classification instead of any GNN models.
Guided by graph properties like sparsity, rank, and feature smoothness, \cite{jin2020graph} presented Pro-GNN which jointly learns clean graph structure and trains robust GNN models together.

\subsection{Summary: Defense on Graph}

From the perspective of defenders, the defense approaches can be designed with or without knowing the specific attacks. Thus, current defense works can be classified into two categories: 1) \textit{Attack-agnostic defenses} are designed to enhance the robustness of graph models against any possible attacks instead of a fixed one. 2) \textit{Attack-oriented defenses} are designed according to the characteristics of specific attacks. The attack-agnostic defenses usually have a wider assumption space of attacks comparing to attack-oriented attack. 
Last, we discuss some future opportunities on adversarial defense in this area.

 \textbf{Attack-agnostic Defense.}
As we summarized in Section~\ref{sec:AD_training}, adversarial training is a typical instance of attack-agnostic defense approach~\cite{feng2019graph, dai2018adversarial, sun2019virtual, xu2019topology, deng2019batch}. It usually generates simple perturbations on graphs or models to train a defense model. In the test phase, some models trained in this way could exhibit good robustness against those perturbations. Some methods~\cite{xu2019topology} trained in this way even attain good defense performance against other specific attacks like Meta-self proposed in~\cite{zugner2019adversarial}. Note that the defense methods are designed and trained without knowing other new attacks.

Besides adversarial training, other works secure the graph model with heuristic assumptions on the attack strategies and outcomes. \cite{tang2020transferring} assumes that there are unpolluted graphs to aid the detection of attacks. \cite{jin2019power, zhu2019robust, ioannidis2020tensor, hou2019alphacyber} propose new GNN architectures to enhance their robustness. \cite{miller2019improving, miller2020topological} directly curates an optimal training set to mitigate the vulnerability of trained models.

\textbf{Attack-oriented Defense.}
Attack-oriented defenses are designed based on the strategy and approach of specific attacks. Namely, the defender has full knowledge of an attack method and the defense method could detect the corresponding attack or curb its performance. Among current defense works, \cite{entezari2020all} first argued the weakness of Nettack~\cite{zugner2018adversarial} and leveraged SVD to defend against Nettack. \cite{jin2019latent} analyzed the strategies and approaches of  Nettack~\cite{zugner2018adversarial} and RL-S2V~\cite{dai2018adversarial} and proposed an adversarial training method. \cite{wu2019adversarial} inspected two gradient-based attacks (i.e., FGSM~\cite{goodfellow2014explaining} and JSMA~\cite{papernot2016limitations}) and applied edge-dropping technique during model training to alleviate the influence of such attacks. Similar to attack-agnostic defenses, some attack-oriented methods exhibit good generability which means it can defend against other unknown attacks. For instance, the defense method proposed in~\cite{wu2019adversarial} could defend the Nettack as well. 
Along with the \textbf{Corresp. Attack} column of Table~\ref{table::defense_summary}, we could see that Nettack and RL-S2V have become benchmark attack methods for defense design and evaluation. 
Some works employ the framework of minimax game~\cite{gaitonde2020adversarial} or optimization~\cite{bojchevski2019certifiable,zugner2019certifiable,jia2020certified} to certify the robustness bounds of graph models under given attacks and defenses. Such kind of defense works are attack-oriented since they have assumed specific attacks.

\textbf{Most Recent Work.} In addition to common tasks such as node classification and link prediction, attacks \mbox{\cite{zhang2020adversarial}} and defenses \mbox{\cite{zhao2021expressive, ren2021integrated, zhou2021robust}} on graph alignment tasks (e.g. graph matching) were also proposed.

 \textbf{Limitations and Future Directions.}
We have been focusing on the contributions of different existing works on graph adversarial defense. Now we summarize some common limitations we observe in this line of research and hint on future directions:
1) Most defense models focus on node-level tasks, especially node classification, while it may be intriguing to shed more light on link- and graph-level tasks like link prediction and graph classification. There is also large potential in more real-life tasks like graph-based search, recommendation, advertisement and etc.
2) While network data are often associated with complex contents nowadays (e.g., timestamps, images, texts), existing defense models have hardly considered the effect of attacks and defenses under the settings of dynamic or other content-rich complex networked systems.
3) Most defense models are relevant to GNNs or GCN in particular, but there are many other graph models and analysis methods, possibly more widely used and less studied (e.g., random walk based models, stochastic block models, and many computational graph properties). How are they sensitive and prone to graph adversarial attacks? Can the improvements in GNN models transfer and generalize to these traditional methods and measures?
4) Most existing works do not study the efficiency and scalability of defense models. As we know, real-world networks can be massive and often frequently evolve, so how to efficiently learn the models and adapt to changes is very important for defenders.
5) While there are standard evaluation protocols and optimization goals for down-stream tasks like node classification and link prediction, defense methods are optimized towards heterogeneous goals like accuracy, robustness, generalizability and so on, and they tend to define their own experimental settings and metrics, rendering fair and comprehensive evaluations challenging.

\section{Metrics} 
\label{sec:metric}

In this section, we summarize the metrics for evaluating attack and defense performance on graph data. We first briefly introduce the general evaluation metrics along with some notes on their specific usage in adversarial performance evaluation. We then give a detailed introduction of particular evaluation metrics designed for attacks and defenses.

\subsection{General Metric}

\subsubsection{Accuracy-based Metric}

According to Table~\ref{table::attack_summary} and Table~\ref{table::defense_summary}, many existing works tackle the node classification problem which is usually a \textit{binary} or \textit{multi-class} classification problem. The accuracy-based metrics like \textbf{Accuracy}, \textbf{Recall}, \textbf{Precision}, and \textbf{F1 score} are all used by existing works to reflect the classification accuracy from different angles. Readers can refer to \cite{wikipedia_matrix} for detailed explanations of those metrics. Note that the \textbf{False Negative Rate} (\textbf{FNR}) and \textbf{False Positive Rate} (\textbf{FPR}) used by~\cite{chen2017practical, wang2019attacking} are two metrics derived from the confusion matrix. FNR is the percentage of false negatives among all actual positive instances, which describes the proportion of positive instances missed by the classifier. Similarly, FPR reflects the proportion of negative instances misclassified by the classifier. \textbf{Adjusted Rand Index} (\textbf{ARI})~\cite{wikipedia_ARI} is an accuracy-based metric without label information. \cite{chen2019multiscale} uses it to measure the similarity between two clusters in a graph.

Besides the above metrics, {Area-under-the-ROC-curve} (\textbf{AUC})~\cite{wikipedia_auc} and \textbf{Average
Precision} (\textbf{AP})~\cite{wikipedia_ap} are widely used, such as by \cite{ioannidis2019graphsac, sun2018data, waniek2018attack, xu2018characterizing, zhu2019robust}. AUC is sensitive to the probability rank of positive instances, which is larger when positive instances are ranked higher than negative instances according to the predicted probability of a classifier. AP is a metric balancing the Precision and Recall where AP is higher when Precision is higher as Recall threshold increase from 0 to 1. Those two metrics could better reflect the classification performance as single scores since they provide an all-around evaluation over the predicted probabilities of all instances.

\subsubsection{Ranking-based Metric}
\textbf{Mean Reciprocal Rank} (\textbf{MRR})~\cite{wikipedia_mrr} and \textbf{Hits@K} are two ranking metrics used by~\cite{pezeshkpour2019investigating, zhang2019towards} to evaluate the performance of link prediction on knowledge graphs. Given a list of items retrieved regarding a query and ranked by their probabilities, the reciprocal rank of a query response is the multiplicative inverse of the rank of the first correct item: 1 for first place, 1⁄2 for second place, 1⁄3 for third place and so on. Hits@K is the number of correct answers among the \textit{top K} items in the ranking list. It can be used to evaluate the performance of recommender system as well~\cite{fang2018poisoning}. \textbf{nDCG@K}~\cite{wikipedia_nDCG} is another metric to evaluate the robustness of recommendation models~\cite{peng2020robust}.

\subsubsection{Graph-based Metric}
The graph-based metrics indicate the specific properties of a graph. \textbf{Normalized Mutual Information} (\textbf{NMI})~\cite{wikipedia_nmi} and \textbf{Modularity}~\cite{wikipedia_modularity} are two metrics used by~\cite{chen2019ga, chen2019multiscale, xuan2019unsupervised} to evaluate the performance of community detection (i.e., clustering) on graphs. NMI is originated from information theory that measures the mutual dependence between two variables. In a community detection scenario, NMI is used to measure the amount of shared information (i.e., similarity) between two communities. Modularity is designed to measure the strength of the division of a graph into clusters. Graphs with high Modularity have dense connections between the nodes within clusters but sparse connections between nodes in different clusters.

\cite{sun2019node} employs a couple of graph property statistics as metrics to evaluate how much the attacker changed the graph (i.e., the imperceptibility of attacks). The metrics include \textbf{Gini Coefficient}, \textbf{Characteristic Path Length}, \textbf{Distribution Entropy}, \textbf{Power Law Exponent}, and \textbf{Triangle Count}. Please refer to~\cite{bojchevski2018netgan} for more details about those metrics. Some more graph statistics metrics include \textbf{Degree Ranking}, \textbf{Closeness Ranking}, \textbf{Betweenness Ranking} used by~\cite{waniek2018hiding} and \textbf{Clustering Coefficient}, \textbf{Shortest Path-length}, \textbf{Diagonal Distance} used by~\cite{xuan2020adversarial}.

\subsection{Adversarial Metric}
Besides the general metrics above, a number of metrics which measure the attack and defense performance on graph data have been proposed or used by existing works. We first present the detailed formulations and descriptions of widely used metrics, and then briefly summarize some unique metrics used by particular papers. The reference after each metric name refers to the first paper that proposes or uses this metric and the references inside the parentheses refer to other attack and defense papers using this metric. 

\subsubsection{Common Metric}

\begin{itemize}

\item \textbf{Attack Success Rate (ASR)~\cite{chen2017practical}}
(\cite{chen2018fast, chen2018link, wang2018attack, bose2019generalizable, ma2019attacking, chen2019time, zang2020graph, takahashi2019indirect, chen2020mga, jin2019latent, zhang2020backdoor, xi2020graph}).
ASR is the most frequently used metric to measure the performance of a giving attack approach:
\begin{equation*}
    \textrm{ASR} = \frac{\textrm{\# Successful attacks}}{\textrm{\# All attacks}}.
\end{equation*}

\item \textbf{Classification Margin (CM)~\cite{zugner2018adversarial}}
(\cite{wu2019adversarial, zhang2019comparing, wang2019adversarial, miller2019improving, zugner2020adversarial}).
CM measures the performance of the integrity attack:
\begin{equation*}
    \textrm{CM}(t) = p_{t, c_{t}} - \max_{c\neq c_{t}}p_{t, c},
\end{equation*}
where $t$ is the target instance, $c_t$ is the
ground-truth class for $t$, $p_{t, c}$ is the probability of $t$ being $c$. The above equation calculates the maximum difference between the probability of ground-truth class and that of other classes. In other words, it shows the extent of an attack flipping the predicted class of a target instance. \cite{miller2019improving} proposed another version of CM:
\begin{equation*}
    \textrm{CM}(t) = \log\frac{p_{t, c_{t}}}{\max_{c\neq c_{t}}p_{t, c}}.
\end{equation*}

When the instance is correctly classified, CM will be positive; otherwise it will be negative.

\item \textbf{Correct/Mis Classification Rate~\cite{bojchevski2019adversarial}}
(\cite{zugner2019adversarial, xu2019topology, entezari2020all, wang2019graphdefense, tang2020adversarial}).
Those two metrics evaluate the attack/defense performance based on the classification results among all instances. 
\begin{align*}
    \textrm{MCR} &= \frac{\textrm{\# Misclassified instances}}{\textrm{\# All instances}}: \\
    \textrm{CCR} &= 1 - \textrm{MCR}.
\end{align*}

\item \textbf{Attacker Budget~\cite{miller2019improving}}
(\cite{dey2019manipulating, miller2020topological}).
Attacker budget is a general metric to measure the minimum perturbations the attacker needs to fulfill its objective. The lower value indicates a better attack performance and a worse defense performance respectively. \cite{dey2019manipulating} takes the number of removed edges as the attacker budget. \cite{miller2019improving, miller2020topological} take the smallest number of perturbations for the attacker to successfully cause the target to be misclassified as the budget.

\item \textbf{Average Modified Links (AML)~\cite{chen2018fast}}
(\cite{chen2018fast, chen2018link, chen2019time, zang2020graph}).
AML is a variance of Adversary budget introduced above. It describes the average number of modified links the attacker needed to meet the attack objective:
\begin{equation*}
    \textrm{AML} = \frac{\textrm{\# Modified links}}{\textrm{\# All attacks}}.
\end{equation*}

\item \textbf{Concealment Measures~\cite{waniek2018hiding}}
(\cite{waniek2018attack, xuan2020adversarial,li2020adversarial}).
The concealment measures are used to evaluate the performance of hiding nodes or communities in a graph~\cite{waniek2018hiding,waniek2018attack,li2020adversarial}. From another perspective, the structural changes introduced by an attack can be used to quantify the concealment of the attack as well~\cite{xuan2020adversarial}.

\item \textbf{Similarity Score~\cite{sun2018data}}
(\cite{zhou2019attacking}).
Similarity score is a general metric to measure the similarity of given instance pairs. It can be used as the goal of integrity attack where the attacker’s goal is either to increase or decrease the similarity
score of a target instance pair. For a node instance in a graph, both of its local structure and node embedding can be used to compute the similarity score.

\end{itemize}

\begin{table*}[hbt!]
\centering
\caption{Summary of datasets (ordered by the frequency of usage within each graph type).}
\label{table::dataset_summary}
\resizebox{\linewidth}{!}{%
\begin{tabular}{l|llllllll}
\toprule
Type & Task & Dataset & Source & \# Nodes  &  \# Edges  &  \# Features  & \# Classes & Paper\\ 
\hline
\multirow{8}{*}{\begin{tabular}[l]{@{}l@{}}Citation \\Network \end{tabular}} 

& Node/Link &  Citeseer & \cite{sen2008collective}  &   3,327     &  4,732  &3,703 &  6  & \begin{tabular}[l]{@{}l@{}l@{}l@{}}  \cite{zugner2018adversarial}, \cite{dai2018adversarial},  \cite{chen2018fast},  \cite{wang2018attack},  \cite{sun2018data}, \cite{bojchevski2019adversarial}, \cite{zugner2019adversarial},  \cite{bose2019generalizable},  \cite{xuan2019unsupervised}, \cite{wu2019adversarial}, \cite{xu2019topology},\\ 
 \cite{chang2020restricted}, \cite{xu2018characterizing}, \cite{feng2019graph}, \cite{zhang2019comparing}, \cite{sun2019virtual}, \cite{chen2019can}, \cite{zhu2019robust}, \cite{wang2019adversarial} , \cite{jin2019power}, \cite{zugner2019certifiable}, \cite{bojchevski2019certifiable},\\ 
\cite{ioannidis2019graphsac},\cite{ioannidis2019edge}, \cite{wang2019graphdefense}, \cite{entezari2020all}, \cite{zang2020graph}, \cite{takahashi2019indirect},  \cite{chen2020mga}, \cite{ioannidis2020tensor}, \cite{dai2019adversarial}, \cite{jin2019latent}, \cite{deng2019batch}, \cite{miller2019improving},\\ 
\cite{miller2020topological},\cite{zugner2020adversarial}, \cite{jin2020graph}, \cite{he2020stealing}, \cite{wang2020scalable}, \cite{zhang2020defensevgae}, \cite{zhang2020gnnguard}, \cite{xu2020edog} \end{tabular}\\
\cline{2-9}

& Node/Link &  Cora & \cite{sen2008collective}  &  2,708  &  5,429  &1,433 & 7  &\begin{tabular}[l]{@{}l@{}l@{}l@{}} \cite{dai2018adversarial},  \cite{chen2018fast},  \cite{wang2018attack},  \cite{sun2018data},  \cite{bojchevski2019adversarial},  \cite{zugner2019adversarial},  \cite{bose2019generalizable},  \cite{xuan2019unsupervised}, \cite{wu2019adversarial},  \cite{xu2019topology}, \cite{chang2020restricted},\\
 \cite{xu2018characterizing}, \cite{feng2019graph}, \cite{zhang2019comparing},  \cite{sun2019virtual}, \cite{chen2019can}, \cite{zhu2019robust}, \cite{wang2019adversarial} , \cite{jin2019power}, \cite{ioannidis2019graphsac}, \cite{ioannidis2019edge}, \cite{wang2019graphdefense},\\ 
\cite{zang2020graph}, \cite{chen2020mga}, \cite{ioannidis2020tensor}, \cite{dai2019adversarial}, \cite{jin2019latent}, \cite{deng2019batch}, \cite{miller2019improving}, \cite{miller2020topological}, \cite{jin2020graph}, \cite{he2020stealing}, \cite{wang2020scalable}, \cite{zhang2020defensevgae},\\
\cite{zhang2020gnnguard}, \cite{xu2020edog}\end{tabular}
\\
\cline{2-9}

& Node &  Pubmed & \cite{sen2008collective}  &  19,717     &  44,338  &500 & 3  & \begin{tabular}[l]{@{}l@{}}\cite{dai2018adversarial},  \cite{zugner2019adversarial}, \cite{chang2020restricted}, \cite{sun2019node}, \cite{sun2019virtual}, \cite{zhu2019robust}, \cite{jin2019power}, \cite{zugner2019certifiable}, \cite{bojchevski2019certifiable}, \cite{ioannidis2019graphsac}, \cite{ioannidis2019edge}, \cite{tang2020transferring},\\ 
\cite{ioannidis2020tensor}, \cite{jin2019latent}, \cite{deng2019batch}, \cite{miller2020topological}, \cite{zugner2020adversarial},\cite{zugner2020adversarial}, \cite{jin2020graph}, \cite{he2020stealing}, \cite{wang2020scalable}\end{tabular}\\
\cline{2-9}

& Node &  Cora-ML &  \cite{mccallum2000automating} &  2,995  &  8,416  &2,879 & 7  &  \cite{zugner2018adversarial}, \cite{zugner2019certifiable}, \cite{bojchevski2019certifiable}, \cite{sun2019node},\cite{entezari2020all}, \cite{takahashi2019indirect}, \cite{jin2019latent}, \cite{zugner2020adversarial} \\
\cline{2-9}

& Node/Community & DBLP & \cite{tang2008arnetminer}  &   -     &   -  & -&  -  & \cite{li2020adversarial}, \cite{jin2019latent}, \cite{jia2020certified}, \cite{wang2020scalable}\\

\hline

\multirow{12}{*}{\begin{tabular}[l]{@{}l@{}}Social \\Network \end{tabular}}

& Node/Link & PolBlogs & \cite{adamic2005political}  &  1,490     &  19,025  & -&  2 & \begin{tabular}[l]{@{}l@{}} \cite{zugner2018adversarial},  \cite{chen2018fast},  \cite{bojchevski2019adversarial}, \cite{wu2019adversarial},  \cite{chen2019multiscale}, \cite{zhang2019comparing}, \cite{chen2019can}, \cite{wang2019adversarial} ,\cite{ioannidis2019graphsac},\cite{ioannidis2019edge},\\
\cite{entezari2020all}, \cite{zang2020graph}, \cite{chen2020mga}, \cite{ioannidis2020tensor}, \cite{jin2019latent}, \cite{miller2020topological}, \cite{jin2020graph}, \cite{zhang2020defensevgae}\end{tabular}\\
\cline{2-9}

& Node/Link &  Facebook  & \cite{leskovec2007graph}   &  -     &  -  &- & -  &  \cite{waniek2018attack},  \cite{chen2018link},  \cite{sun2018data},  \cite{zhou2019attacking},  \cite{wang2019attacking}\\
\cline{2-9}

& Node/Community &  Google+  &  \cite{leskovec2007graph}  &  107,614     &  13,673,453  &- & -  &  \cite{waniek2018hiding}, \cite{waniek2018attack}, \cite{wang2019attacking}\\
\cline{2-9}

& Node &  Reddit & \cite{hamilton2017inductive}  & 1,490       &  19,090  &300 & 2  & \cite{wang2019graphdefense}, \cite{tang2020transferring}, \cite{wang2020scalable}\\
\cline{2-9}

& Community &  Dolphin & \cite{lusseau2003bottlenose}  &  62     &  159  &- & -  &  \cite{chen2019ga},  \cite{xuan2019unsupervised},\cite{yu2019target}\\
\cline{2-9}

& Community & WTC 9/11 & \cite{krebs2002mapping}  &    36    &   64 & -&  -  & \cite{waniek2018hiding}, \cite{waniek2018attack} \\
\cline{2-9}

& Community & Email & \cite{leskovec2007graph}  &   1,005     & 25,571   &- &  -  & \cite{chen2019multiscale}, \cite{jia2020certified}\\
\cline{2-9}

& Community &  Karate & \cite{zachary1977information}  &  34     &  78  &- & -  &  \cite{chen2019ga},  \cite{xuan2019unsupervised}\\
\cline{2-9}

& Community & Football  & \cite{girvan2002community}  &  115     &  613  &- & -  &  \cite{chen2019ga},  \cite{chen2019multiscale}\\
\cline{2-9}
& Fraud Detection & Yelp & \cite{rayana2015collective}  &   -      &  -    & - &  -  & \cite{dou2020robust}, \cite{zhang2020gcn}\\
\cline{2-9}

& Recommendation & MovieLens & \cite{movielens}  &   -      &  -    & - &  -  & \cite{peng2020robust}, \cite{fang2018poisoning}\\

\hline

\multirow{2}{*}{\begin{tabular}[l]{@{}l@{}}Knowledge \\Graph \end{tabular}}

& Fact/Link &  WN18 & \cite{bordes2013translating}  &  -     &  -  &- & -  & \cite{zhang2019towards}, \cite{pezeshkpour2019investigating}\\
\cline{2-9}

& Fact & FB15k & \cite{bordes2013translating}  &   -     &  -  &- &  -  & \cite{zhang2019towards}\\
\hline

\multirow{5}{*}{Others}

& Node & Scale-free &  \cite{barabasi1999emergence} &    -    &   - &- &  -  & \cite{waniek2018hiding}, \cite{waniek2018attack}, \cite{xuan2020adversarial}\\
\cline{2-9}

& Node & NELL & \cite{yang2016revisiting}  &   65,755      &  266,144    & 5,414 &  210  & \cite{feng2019graph}, \cite{deng2019batch}\\
\cline{2-9}

& Graph & Bitcoin & \cite{weber2019anti}  &   -      &  -    & - &  -  & \cite{xi2020graph}, \cite{zhang2020backdoor}\\
\cline{2-9}

& Graph/Node & AIDS & \cite{riesen2008iam}  &   -    &  -    & - &  -  & \cite{tang2020adversarial}, \cite{xi2020graph}, \cite{he2020stealing}\\
\cline{2-9}

& Graph/Node & DHFR & \cite{sutherland2003spline}  &   -    &  -    & - &  -  & \cite{tang2020adversarial}, \cite{he2020stealing}\\

\bottomrule
\end{tabular}
}
\end{table*}

\subsubsection{Unique Metric}

\begin{itemize}

\item \textbf{Averaged Worst-case Margin (AWM)~\cite{bojchevski2019certifiable}.}
The worst-case margin is the minimum value of the classification margin defined above. The averaged worse-case margin means the value is averaged across a worst-case margin of each batch of data.

\item \textbf{Robustness Merit (RM) \cite{jin2019power}.}
RM is the difference between the post-attack accuracy of the proposed method and the post-attack accuracy of the vanilla GCN model. A greater value indicates a better defense performance.

\item \textbf{Attack Deterioration (AD) \cite{jin2019power}.}
AD is the ratio of decreased amount of accuracy after an attack to the accuracy without attack.

\item \textbf{Average Defense Rate (ADR) \cite{chen2019can}.}
ADR is a metric evaluating the defense performance according to the ASR defined above. It compares the ASR after attacks with or without applying the defense approach.

\item \textbf{Average Confidence Different (ACD) \cite{chen2019can}.}
ACD is a metric evaluating the defense performance based on the average difference between the classification margin after and before the attack of a set of nodes. Such a set of nodes includes correctly classified nodes before the attack.

\item \textbf{Damage Prevention Ratio (DPR) \cite{zhou2019adversarial}.}
Damage prevention measures the amount of damage that can be prevented by the defense. Let $L_{0}$ be the
defender’s accumulated loss when there is no attack. Let $L_{A}$ be the defender’s loss under some attack $A$ when the defender cannot make any reliable queries. $L_{D}$ denotes the
loss when the defender make reliable queries according to
a certain defense strategy $D$. DPR can be defined as follows:
\begin{equation*}
    \textrm{DPR}_{A}^{D} = \frac{L_{A}-L_{D}}{L_{A}-L_{0}}.
\end{equation*}

\item \textbf{Certified Accuracy~\cite{jia2020certified}.}
It is proposed to evaluate the certification method for robust community detection models against adversarial attacks.
The certified accuracy $CK(l)$ is the fraction of sets of victim nodes that proposed method can provably detect as in the same community when an attacker adds or removes at most $l$ edges in the graph.

\item \textbf{Practical Effect~\cite{dou2020robust}.}
Since the attacker may target at practical effect of attacks like boosting item revenue or reputation, 
\cite{dou2020robust} proposed a revenue-based metric to measure the performance of attacks and defenses from a practical angle.

\end{itemize}
\section{Dataset and Application} \label{sec:dataset}

Table~\ref{table::dataset_summary} summarizes some common datasets used in adversarial attack and defense works on graph data.
The first four citation graphs have been widely used as node classification benchmarks in previous work~\cite{kipf2016semi, velivckovic2017graph, velivckovic2018deep, wu2019simplifying}.
\cite{sun2018data} also studies the adversarial link prediction problem on Cora and Citeseer.
DBLP includes multiple citation datasets with more metadata information.
Thus it can be used to study the community detection task~\cite{jia2020certified}.
Among the social network datasets, PolBlogs is another dataset used especially in adversarial settings where blogs are nodes and their cross-references are edges.
Reddit and Facebook are two larger graph datasets compared to citation datasets.
Since there are multiple versions of Facebook datasets used across different papers, we omit its statistics.
WTC 9/11, Email, Dolphin, Karate, and Football are five benchmark datasets for community detection.
Some recent works also studied attacks and defenses of recommender system~\cite{peng2020robust,fang2018poisoning} and review system~\cite{dou2020robust, zhang2020gcn} based on the Yelp and MovieLens data.
\cite{pezeshkpour2019investigating, zhang2019towards} investigated the adversarial attacks and defenses on knowledge graphs using two knowledge graph benchmarks WN18 and FB15k.
Scale-free network is a typical type of graph synthesized by graph generation models. 
Some works also employ other graph generation models like Erdős-Rényi model to generate graphs to facilitate their experiments~\cite{waniek2018hiding,zhou2019attacking,waniek2018attack,chen2020network, kenlay2020stability, breuer2020friend, zhang2020backdoor}. 
Besides the node-level tasks, Bitcoin, AIDS, and DHFR datasets which contain multiple graphs are used to investiagte the robustness of graph classfication models~\cite{zhang2020backdoor,xi2020graph,tang2020adversarial,he2020stealing}.
Among them, Bitcoin is a Bitcoin transaction dataset, 
AIDS contains biological graphs to represent the antiviral character of different biology compounds,
and DHFR contains graphs to represent the chemical bond type. 

\begin{table*}
\centering
\caption{Summary of open-source implementations of algorithms.}
\label{table::code_summary}
\resizebox{\linewidth}{!}{%
\begin{tabular}{l|cll}
\toprule
Type & Paper &Algorithm  & Link \\ 
\hline

\multirow{12}{*}{Graph Attack} 

& \cite{chen2018fast}  & FGA    & \url{https://github.com/DSE-MSU/DeepRobust}\\

&  \cite{waniek2018hiding} &  DICE   & \url{https://github.com/DSE-MSU/DeepRobust}\\

& \cite{zugner2018adversarial}  & Nettack    & \url{https://github.com/danielzuegner/nettack}\\

&\cite{dai2018adversarial}  & RL-S2V, GraArgmax  & \url{https://github.com/Hanjun-Dai/graph_adversarial_attack}\\

& \cite{wu2019adversarial}  & IG-Attack    & \url{https://github.com/DSE-MSU/DeepRobust}\\

& \cite{zugner2019adversarial} &  Meta-self, Greedy &\url{https://github.com/danielzuegner/gnn-meta-attack}\\

& \cite{bojchevski2019adversarial} &    ICML-19   & \url{https://github.com/abojchevski/node_embedding_attack}\\

& \cite{xu2019topology} &  PGD, Min-max     & \url{https://github.com/KaidiXu/GCN_ADV_Train}\\

& \cite{chang2020restricted} &   GF-Attack  & \url{https://github.com/SwiftieH/GFAttack}\\

& \cite{sun2019node}  & NIPA    & \url{https://github.com/DSE-MSU/DeepRobust}\\

& \cite{zang2020graph} &   GUA  & \url{https://github.com/chisam0217/Graph-Universal-Attack}\\

& \cite{dou2020robust} &  IncBP, IncDS  & \url{https://github.com/YingtongDou/Nash-Detect}\\

\hline
\multirow{14}{*}{Graph Defense}

& \cite{feng2019graph}  &  GraphAT & \url{https://github.com/fulifeng/GraphAT}\\

& \cite{dai2019adversarial} &  AdvT4NE  &\url{https://github.com/wonniu/AdvT4NE_WWW2019}\\

& \cite{zhu2019robust} &  RGCN  &\url{https://github.com/DSE-MSU/DeepRobust}\\

& \cite{wu2019adversarial} &   GCN-Jaccard   & \url{https://github.com/DSE-MSU/DeepRobust}\\

& \cite{xu2019topology}  &  Adversarial
Training  &\url{https://github.com/KaidiXu/GCN_ADV_Train}\\

& \cite{zugner2019certifiable}&  Robust-GCN  & \url{https://github.com/danielzuegner/robust-gcn}\\

& \cite{tang2020transferring} & PA-GNN   & \url{https://github.com/tangxianfeng/PA-GNN}\\

& \cite{jin2019power} &  r-GCN, VPN  & \url{https://www.dropbox.com/sh/p36pzx1ock2iamo/AABEr7FtM5nqwC4i9nICLIsta?dl=0}\\

& \cite{bojchevski2019certifiable} &   Graph-cert   & \url{https://github.com/abojchevski/graph_cert}\\

& \cite{entezari2020all} &  GCN-SVD  & \url{https://github.com/DSE-MSU/DeepRobust}\\

& \cite{jin2020graph} &  Pro-GNN  & \url{https://github.com/DSE-MSU/DeepRobust}\\

& \cite{zhang2020defensevgae} &  DefenseVGAE  & \url{https://github.com/zhangao520/defense-vgae}\\

& \cite{kenlay2020stability} &  SPGF  & \url{https://github.com/henrykenlay/spgf}\\

& \cite{dou2020robust} &  Nash-Detect  & \url{https://github.com/YingtongDou/Nash-Detect}\\

\hline
\multirow{4}{*}{Other Baseline}

& \cite{goodfellow2014explaining} &   FGSM    & \url{https://github.com/1Konny/FGSM}\\

& \cite{papernot2016limitations} &    JSMA    & \url{https://github.com/tensorflow/cleverhans}\\

& \cite{biggio2013evasion} &   Gradient Attack (GA)     & \url{https://github.com/bethgelab/foolbox/blob/master/foolbox/attacks/gradient.py}\\

& \cite{finn2017model} &   First-order     & \url{https://github.com/cbfinn/maml}\\

\hline
\multirow{1}{*}{Benchmark}

& \cite{zheng2021graph} & GRB & \url{https://github.com/THUDM/grb}\\

\bottomrule
\end{tabular}}
\end{table*}

\textbf{Future Directions.} 
Besides the datasets listed in Table~\ref{table::dataset_summary}, it is worth noting some other datasets which get less attention but could be studied in future research. To the best of our knowledge, \cite{hou2019alphacyber} is the first and only paper to examine the vulnerability of Heterogeneous Information Network (HIN) which is a graph model with heterogeneous node and edge types~\cite{shi2016survey}. Though HIN has been applied to many security applications like malicious user detection~\cite{zhang2019key}, spam detection~\cite{li2019spam}, and financial fraud detection~\cite{hu2019cash}, its robustness against adversarial attacks remain largely unexplored. A recent study~\cite{gaitonde2020adversarial} firstly gives a formulation of adversarial attacks on opinion propagation on graphs with a spectral form that could be used to study the opinion dynamics of social network. \cite{zhang2019towards, pezeshkpour2019investigating} are the first two works studying the adversarial attacks and defenses on Knowledge Graph (KG) models. As the research of KG becomes popular in recent years, its security issue needs to be noticed as well. The security of dynamic graph models~\cite{chen2019time} is another avenue of research as well.

Besides the above works and datasets, there has been little discussion on the security issues of many other graph types and their related applications. To name a few, the biology graph, causal graph, and bipartite graph have attracted significant research attention but few work has studied potential attacks and their countermeasures on those graphs. From the perspective of applications, as the GNNs having been successfully applied to recommender system, computer vision and natural language processing~\cite{wu2019comprehensive}, adversarial attacks and defenses on graph data under those specific applications is another promising research direction with de facto impacts.



\section{Conclusion}
In this work, we cover the most released papers about adversarial attack and defense on graph data as we know them today.
We first give a unified problem formulation for adversarial learning on graph data,
and give definitions and taxonomies to categorize the literature on several levels.
Next, we summarize most existing imperceptible perturbations evaluation metrics, datasets
and discuss several principles about imperceptibility metric.
Then, we analyze the contributions and limitations of existing works.
Finally, we outline promising future research directions and opportunities that may come from this effort.

\section*{Acknowledgement}
Thanks to Yixin Liu's (yila22@lehigh.edu) contributions to this work and the future maintenance of this work. This work is also supported in part by NSF under grants III-1763325, III-1909323,  III-2106758, and SaTC-1930941. 

\appendices
\section{Open-source resources}

In this work, we not only develop taxonomies for all relevant works based on different criteria, but also summarize the corresponding datasets and metrics that are frequently used.
Moreover, in Table \ref{table::code_summary}, we also provide links to the open-source implementation of popular methods.
To date, it has been a benchmark, Graph Robustness Benchmark (GRB) \mbox{\cite{zheng2021graph}}, that aims to provide a standardized evaluation framework for measuring adversarial attacks and defenses on the node classification task. We hope our work can facilitate the community towards the construction of benchmarks of other graph tasks.

\section{Comparison of existing surveys and ours} 

We conducted the first comprehensive review on graph adversarial attack and defense, and we established a unified formulation for adversarial learning on graph data, which is also utilized in the following two surveys \mbox{\cite{jin2020adversarial, chen2020survey}}. In contrast to our work, Jin et al., \mbox{\cite{jin2020adversarial}} provide implementation details and the specified mathematical modeling of representative algorithms instead of unified formulation, and they conduct empirical studies on graph adversarial attacks. However, the survey lacks introduction to evaluation and perturbation metrics, which are significant to measure the quality of attacks and defenses on graphs. In addition, limitations of current work and future directions are not fully discussed. Although \mbox{\cite{chen2020survey}} and ours have similar structure and content, we provide clearer taxonomies and more comprehensive analysis of attack/defense methods. We also summarize specific strategies of each paper. Besides, our survey covers more papers including most recent work compared with the other two reviews.




\bibliographystyle{plain}
\bibliography{reference.bib}
\vspace{-1cm}
\begin{IEEEbiography}[{\includegraphics[width=1in,height=1.5in, clip, keepaspectratio]{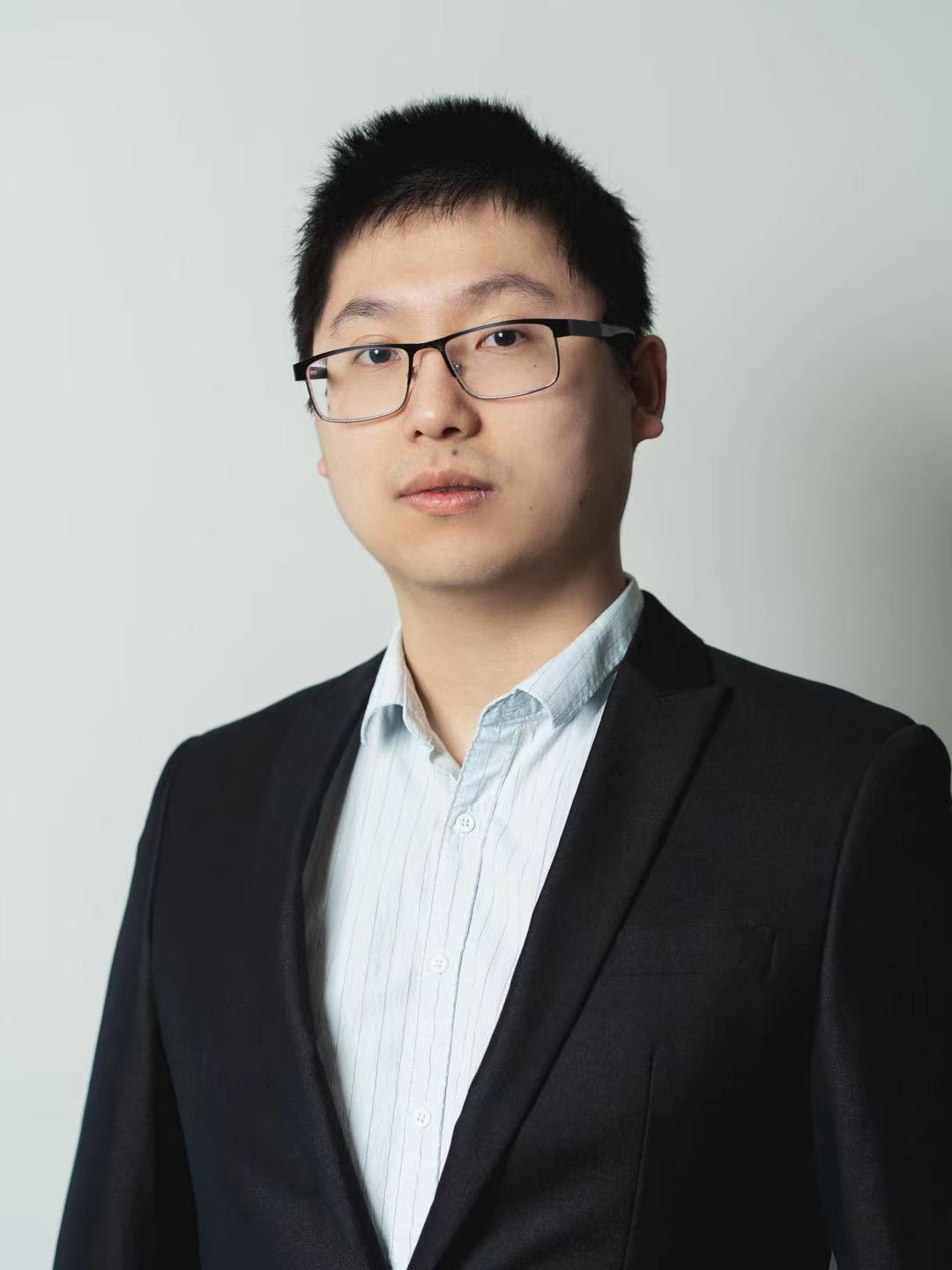}}] {Lichao Sun} is currently an Assistant Professor in the Department of Computer Science and Engineering at Lehigh University. Before that, he received my Ph.D. degree in Computer Science at University of Illinois, Chicago in 2020, under the supervision of Prof. Philip S. Yu. Further before, he obtained M.S. and B.S. from University of Nebraska Lincoln. His research interests include security and privacy in deep learning and data mining. He mainly focuses on AI security and privacy, social networks, and natural language processing applications. He has published more than 45 research articles in top conferences and journals like CCS, USENIX-Security, NeurIPS, KDD, ICLR, AAAI, IJCAI, ACL, NAACL, TII, TNNLS, TMC.
\end{IEEEbiography}
\vspace{-1cm}

\begin{IEEEbiography}[{\includegraphics[width=1in,height=1.5in, clip, keepaspectratio]{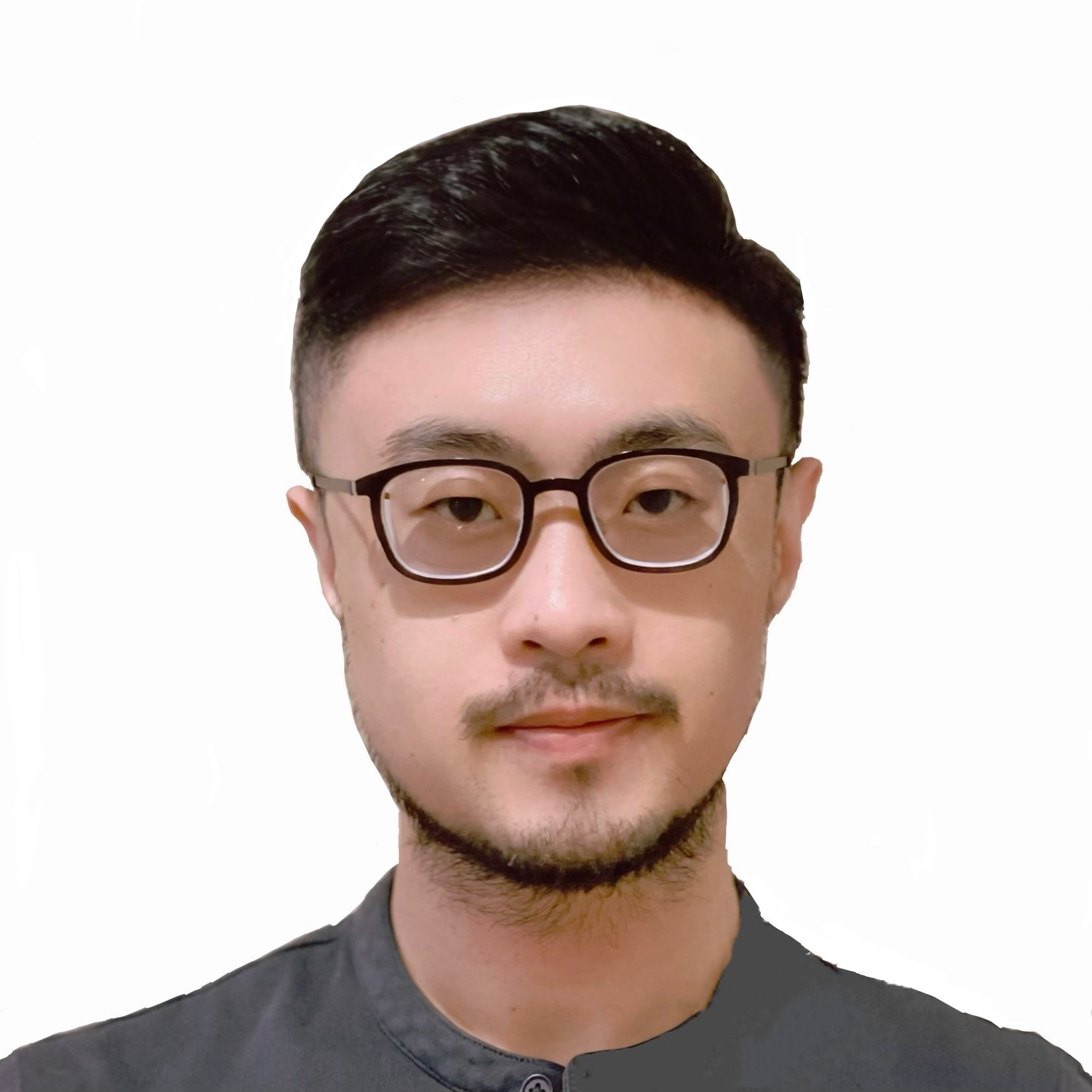}}] {Yingtong Dou} received the B.E. degree in IoT engineering from Beijing University of Posts and Telecommunications, Beijing, China, in 2017. He is currently pursuing the Ph.D. degree in computer science at University of Illinois at Chicago, Chicago, IL, USA.
He has published several papers in top-tier conferences including KDD, WWW, SIGIR, and CIKM. His research interest includes graph mining, fraud detection, social media analysis, and machine learning security. 
\end{IEEEbiography}
\vspace{-1cm}

\begin{IEEEbiography}[{\includegraphics[width=1in,
height=1.25in,clip,keepaspectratio]{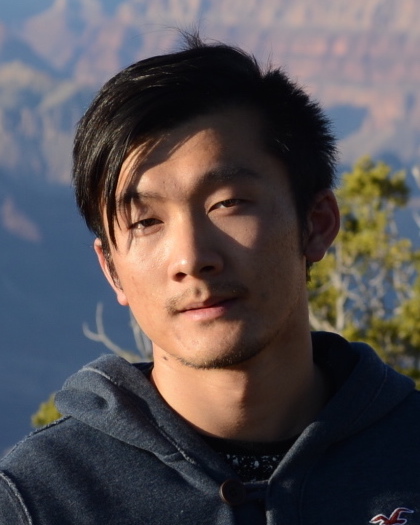}}]{Carl Yang} is an Assistant Professor in Emory University. He received his Ph.D. in Computer Science at University of Illinois, Urbana-Champaign in 2020, and B.Eng. in Computer Science and Engineering at Zhejiang University in 2014. His research interests span graph data mining, applied machine learning, knowledge graphs and federated learning, with applications in recommender systems, biomedical informatics,  neuroscience and healthcare. Carl's research results have been published in top venues like TKDE, KDD, WWW, NeurIPS, ICML, ICLR, ICDE, SIGIR and ICDM. He also received the Dissertation Completion Fellowship of UIUC in 2020, the Best Paper Award of ICDM in 2020, the Dissertation award finalist of KDD in 2021, and the Best Paper Award of KDD Health Day in 2022.
\end{IEEEbiography}
\vspace{-1cm}

\begin{IEEEbiography}[{\includegraphics[width=1in,height=1.5in, clip, keepaspectratio]{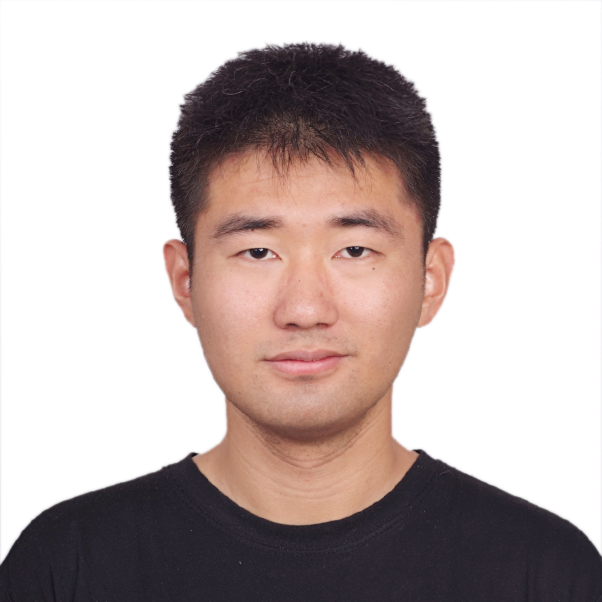}}]{Kai Zhang} is a Ph.D. student at Lehigh University, US. His research interests include machine learning and data mining. He recently focuses on federated learning, security \& data privacy in machine learning, and Edge AI. Before joining Lehigh, he was a Reearch Assistatn at Embry-Riddle Aeronautical University working on large-scale flight dispatching for disaster evacuation with big aviation data. The system design has been featured by the Annual National Mobility Summit and the National Transportation Library. 

\end{IEEEbiography}
\vspace{-1cm}

\begin{IEEEbiography}[{\includegraphics[width=1in,height=1.5in, clip, keepaspectratio]{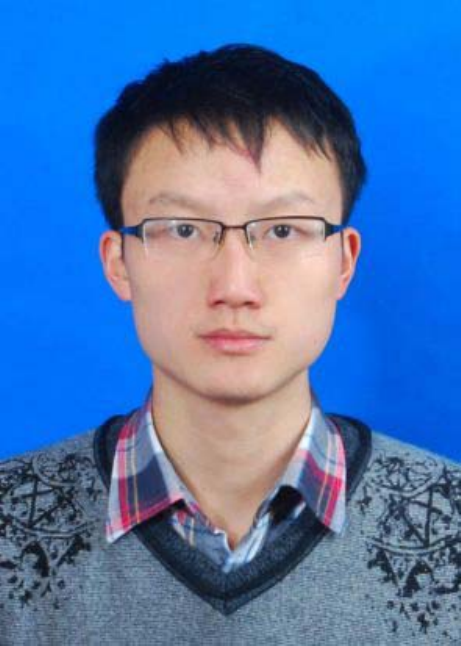}}] {Ji Wang} received the PhD degree in information system from National University of Defense Technology, Changsha, China, in 2019. He is currently an Assistant Professor at the College of Systems Engineering, National University of Defense Technology. His research interests include deep learning and edge intelligence. He has published more than 20 research articles in refereed journals and conference proceedings such as IEEE TC, IEEE TPDS, SIGKDD, and AAAI. He was a visiting PhD student at University of Illinois at Chicago from March 2017 to September 2018 under the supervision of Prof. Philip S. Yu.
\end{IEEEbiography}
\vspace{-1cm}

\begin{IEEEbiography}[{\includegraphics[width=1in,height=1.5in, clip, keepaspectratio]{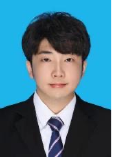}}] {Yixin Liu} received his BEng degree from the Department of Software, South China University of Technology, in 2022. He is currently pursuing a Ph.D. degree in the Department of Computer Science and Engineering at Lehigh University. His research interests mainly include machine learning security, trust-worthy machine learning and adversarial learning.
\end{IEEEbiography}
\vspace{-1cm}

\begin{IEEEbiography}[{\includegraphics[width=1in,height=1.5in, clip, keepaspectratio]{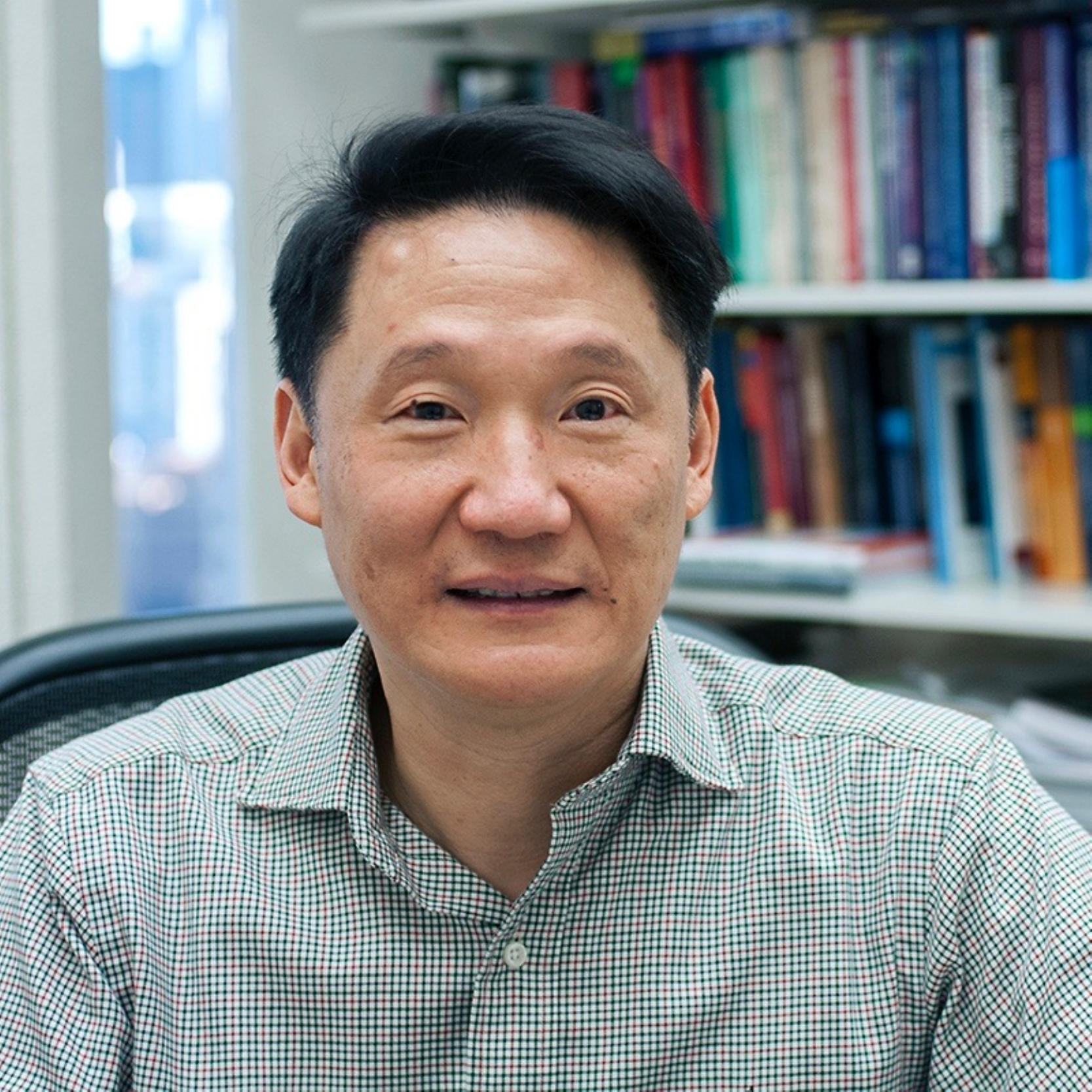}}] {Philip S. Yu} received the B.S. Degree in E.E. from National Taiwan University, the M.S. and Ph.D. degrees in E.E. from Stanford University, and the M.B.A. degree from New York University. He is a Distinguished Professor in Computer Science at the University of Illinois at Chicago and also holds the Wexler Chair in Information Technology.   Before joining UIC, Dr. Yu was with IBM, where he was manager of the Software Tools and Techniques department at the Watson Research Center. His research interest is on big data, including data mining, data stream, database and privacy. He has published more than 1,200 papers in refereed journals and conferences. He holds or has applied for more than 300 US patents. Dr. Yu is a Fellow of the ACM and the IEEE. Dr. Yu is the recipient of ACM SIGKDD 2016 Innovation Award for his influential research and scientific contributions on mining, fusion and anonymization of big data. 
He also received the ICDM 2013 10-year Highest-Impact Paper Award, and the EDBT Test of Time Award (2014). He was the Editor-in-Chiefs of ACM Transactions on Knowledge Discovery from Data (2011-2017) and IEEE Transactions on Knowledge and Data Engineering (2001-2004).
\end{IEEEbiography}
\vspace{-1cm}

\begin{IEEEbiography}[{\includegraphics[width=1in,height=1.2in,clip,keepaspectratio]{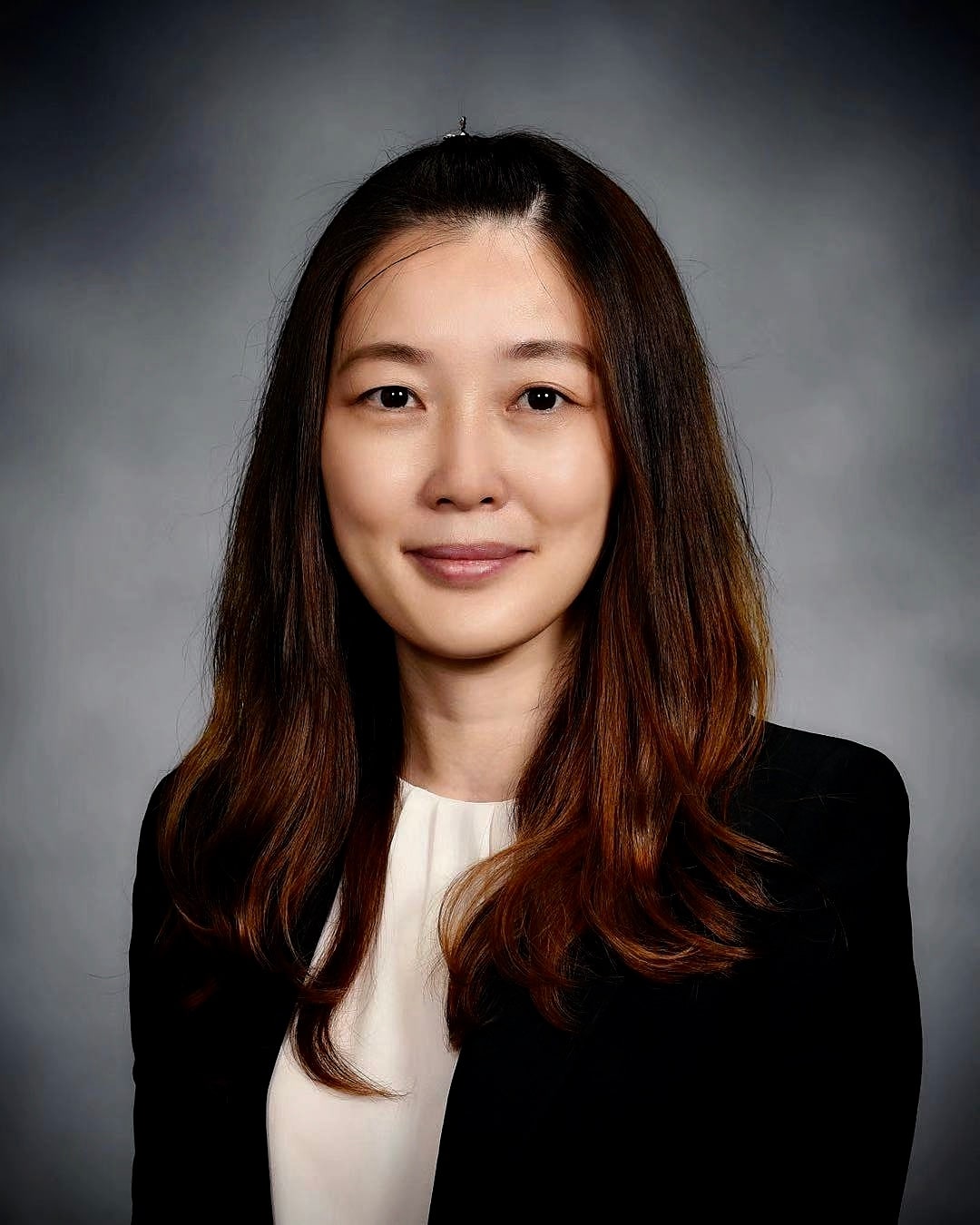}}]
{Lifang He} is currently an Assistant Professor in the Department of Computer Science and Engineering at Lehigh University. Before her current position, Dr. He worked as a postdoctoral researcher in the Department of Biostatistics and Epidemiology at the University of Pennsylvania. Her current research interests include machine learning, data mining, tensor analysis, with major applications in biomedical data and neuroscience. She has published more than 100 papers in refereed journals and conferences, such as NIPS, ICML, KDD, CVPR, WWW, IJCAI, AAAI, TKDE, TIP and AMIA.  
\end{IEEEbiography}
\vspace{-1cm}

\begin{IEEEbiography}[{\includegraphics[width=1in,height=1.5in, clip, keepaspectratio]{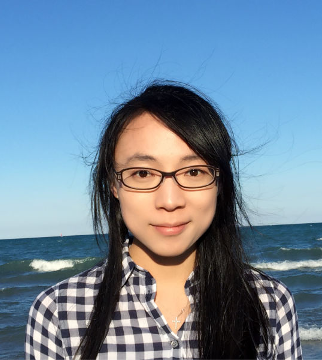}}]{Bo Li} is an assistant professor in the department of Computer Science at University of Illinois at Urbana–Champaign, and is a recipient of the Symantec Research Labs Fellowship, Rising Stars, MIT TR 35, and best paper awards in several machine learning and security conferences. Previously she was a postdoctoral researcher in UC Berkeley. Her research focuses on both theoretical and practical aspects of security, machine learning, privacy, game theory, and adversarial machine learning. She has designed several robust learning algorithms, a scalable framework for achieving robustness for a range of learning methods, and a privacy preserving data publishing system. Her work have been featured by major publications and media outlets such as Nature, Wired, Fortune, and IEEE Spectrum.
\end{IEEEbiography}

\end{document}